\documentclass[3p,numbers,sort&compress]{elsarticle}
\usepackage{amsmath,amssymb,graphicx,color,here,subfig,natbib,upgreek,hyperref}

\providecommand{\bibinfo}[2]{#2}
\providecommand{\doi}[1]{doi:~#1}

\begin{document}
	
	\begin{frontmatter}
		
		\title{Reduced-order modeling of squeeze-film damped electrostatically actuated microbeams using complex modes of the coupled problem}
		
		\author[els]{Fehmi NAJAR \corref{cor1}}
		\cortext[cor1]{Corresponding author, e-mail: f.najar@psau.edu.sa}
		\address[els]{Department of Mechanical Engineering, College of Engineering, \\
			Prince Sattam bin Abdulaziz University, Al-Kharj 16273, Saudi Arabia.}
		
		\begin{abstract}
			A reduced-order model is developed for the nonlinear dynamics of electrostatically actuated microbeams, including the full squeeze film effect. The model uses Petrov--Galerkin projection on damped complex modes of the fully coupled fluid--structure problem. The Euler--Bernoulli theory for calmped--clamped and clamped--free microbeams, with mid-plane stretching and electrostatic nonlinearities, and the nonlinear compressible Reynolds equation are discretized by the differential quadrature method for the static solution and eigenvalue problem. The calculated eigenmodes are either complex conjugates describing the coupled behavior of the device or real depicting the fluid relaxation. The natural frequencies and quality factors are compared to experimental and numerical results found in the literature. Depending on the ambient pressure, the geometry of the microbeam, and the fluid boundary conditions, squeeze-film effect can impact not only the effective damping of the device but also its dynamics. It is shown that in some cases dropping fluid relaxation modes cannot be compensated by any set of purely structural modes. The non-classical damping introduced by the fluid modes is a spatially distributed phase difference between pressure and deflection. The convergence analysis of the dynamic response resulting from the reduced-order model showed that a basis with one set of coupled modes and one fluid mode can reproduce the results of a fully discretized model within 1.5\% error and at much faster computational time.
		\end{abstract}
		
		
	\end{frontmatter}
	
	%%%%%%%%%%%%%%%%%%%%%%%%%%%%%%%%%%%%%%%%%%%%%%%%%%%%%%%%%%%%%%%%%%%%%%%%%%%%%%%%%%%%%%%%%%%%%%%%%%%%%%%%%%%%%%%%%%%%%%%%%%%%%
	\section{Introduction}
	Electrostatically actuated microbeams are widely used as resonant sensors and filters to switches and energy harvesters in microelectromechanical systems (MEMS). In these devices the movable electrode is separated from the substrate by a few micrometers thick air film, and the motion of the structure squeezes the gas in and out of the electrostatic gap. The resulting squeeze-film effect is similar to a viscous damper, it is usually the dominant mechanism of energy dissipation especially for law pressure or low actuation frequencies \citep{Griffin1966,blech1983isothermal}. However, this mechanism is not purely dissipative because of the compressibility of the gas, the air film can also act as a spring. At high frequencies or when the fluid boundary condition are weakly vented, the spring effect can shift the resonance frequency and introduce a hardening effect \citep{Dolleman2021}. The relative weight of the viscous and elastic contributions depends on the squeeze number, on the fluid boundary conditions, and on air rarefaction, which alters the effective viscosity of the gas at reduced pressure \citep{Veijola1995,sumali2007squeeze}.
	
	When air-fil gap is much smaller than the electrodes dimensions of the MEMS device, its pressure is governed by the compressible Reynolds equation \citep{Griffin1966,blech1983isothermal,hamrock2004fundamentals}, which couples a two-dimensional pressure field to the deflection of the flexible electrode \citep{najar2020multifidelity,najar2020double}. Simplified models used approximate coupling by only taking into account the linear viscous damping effect with coefficients obtained from analytical solutions for rigid parallel plates \citep{Khaliq2005,Huang2014}. Such models are known to overestimate the damping of flexible structures and to misrepresent the response when the deflection is not uniform \citep{pandey2007effect}.
	
	Other approaches retain the fluid coupling with the flexural modes. \citet{nayfeh2004squeeze} used perturbation methods to express the pressure analytically in terms of the structural mode shapes, which reduces the dimension of the coupled problem considerably, and extended the approach to microplates under large electrostatic loads \citep{younis2007simulation}. Static deflection due to DC voltage bias was incorporated by \citet{li2007new}. They coupled the compressible Reynolds and microbeam equations for clamped--free and clamped--clamped boundary conditions, representing the pressure as the product of a parabolic profile across the width and a cosine series along the length of the microbeam. Similarly, Chaterjee and Pohit developed a semi-analytical model of clamped--free microbeam, with the Euler--Bernoulli theory and the linearized compressible Reynolds equation for small-amplitude motion about the statically deflected position \citep{Chaterjee_Pohit_2010}. They compared their results with coupled-field finite element simulations. Taking into account the dependence of the effective viscosity on the local gap through a variable Knudsen number, they extended the analysis to the first three flexural modes and showed that the influence of the DC bias on the quality factor and the resonance frequency differs from one mode to another \citep{chaterjee2012squeeze}. These works focus on the linearized damping characteristics about the static operating point, however, the model of \citet{li2007new} assumed spatial form of the pressure by the product of an assumed parabolic function along the width and a cosine series along the length. The results are validated by finite element and experimental data under DC bias and a transient response.
	
	When the fully coupled problem is solved, the fluid domain must be discretized alongside the structure. \citet{Ostasevicius2008} used the finite element software Comsol to implement the nonlinear compressible Reynolds equation with a cantilever beam. They concluded that squeeze-film damping can increase the natural frequency of the device. \citet{younis2007simulation} developed a finite element model to estimate natural frequencies, mode shapes and quality factor using a von Karman plate model under a DC bias and accounting for squeeze film damping through the nonlinear Reynolds equation. They used the singular perturbation technique, previously developed by \citet{nayfeh2004squeeze}, to implement a one-way coupling of the pressure distribution. They concluded that using a particular value for the quality factor when modeling microbeams under electrostatic forcing might lead to erroneous results.
	
	Using the differential quadrature method (DQM) to discretize the nonlinear coupled Euler-Bernoulli and Reynolds equations under DC and AC electrostatic forcing under different configuration and then integrating the Galerkin decomposition for the structural equation, Najar and co-worker \citep{sassi2018square,najar2020double,najar2020multifidelity} derived the forced response of MEMS structures and obtained the frequency response curves. The simulated solutions were compared with other numerical and experimental results. \citet{Ghommem2020} extended the same model by incorporating the fluid inertial effect for liquid environments through an added mass coefficient. \citet{jallouli2019modeling} modeled capacitive circular microplates using the von K\'arm\'an plate theory and the nonlinear Reynolds equation with a size-corrected effective viscosity, discretizing both domains by the differential quadrature method. The linear effects of the film were obtained from the complex eigenvalues of the coupled problem, which showed that the air film adds both stiffness and damping to the microplate, and the model was validated against finite element simulations and measurements. The nonlinear response showed that an increase of the actuation pressure enlarges the bistability domain.
	
	The dimension of the coupled problem remains the main difficulty especially for dynamic analysis and in the frequency domain. Reduced-order models reviewed by \citet{nayfeh2005reduced}, reported that in several approaches the deflection is expanded on the undamped structural mode shapes and the Galerkin method yields a small set of modal equations. The squeeze film is represented by damping coefficients projected onto these modes or solved separately. Modal projection has also been applied to the Reynolds equation itself, after transforming it into a form suitable to a structural basis \citep{missoffe2008reduced}. Snapshot-based techniques, such as the proper orthogonal decomposition, construct the basis a posteriori from simulations of the full model \citep{liang2002proper,gobat2022reduced}. The method requires a full model simulation to be run first, and the extracted basis is limited to the conditions under which they were obtained. In all these approaches the basis is built from the structure alone or from precomputed trajectories, and the dynamics proper to the gas film enter only through the projection.
	
	By examining the structural and fluid governing equations, one can remark that the pressure directly loads the microbeam equation of motion, however, the pressure equation is driven by the velocity of the structure. This asymmetric coupled system is non-self-adjoint. Its eigenvectors are complex, and the problem is of the non-classically damped type. The response can be decoupled by complex modal superposition in the state space, using the right and left eigenvectors of the system \citep{meirovitch2010fundamentals,gruber2018dual}. Complex modes introduce relative phase between the degrees of freedom \citep{hoen2005engineering}, and when a reduced set of them is retained, the choice of the modes is not dictated by frequency alone \citep{dedomenico2019dynamic}. Over-critically damped, non-oscillatory modes in particular may contribute significantly to the response even though they have no counterpart in the undamped problem \citep{chen2020generalized}.
	
	To the best of the author's knowledge, the complex modes of the fully coupled fluid--structure problem have not been used as the basis of a reduced-order model for the nonlinear dynamics of squeeze-film-damped microbeams, and the role played in such a basis by the non-oscillatory modes of the film has not been examined.
	
	The present work addresses this question. The coupled problem formed by the nonlinear Euler--Bernoulli equation of an electrostatically actuated microbeam and the compressible nonlinear Reynolds equation of the air film is first used to generate complex eigenvectors by discretizing the partial differential equations using DQM around the static deflection under a DC bias. The calculated left and right eigenvectors, made biorthonormal, define a Petrov--Galerkin projection that yields a reduced model directly from the physics of the coupled problem, without precomputed simulations. The eigenvectors consist of two sets of modes of distinct nature: complex-conjugate pairs describing coupled oscillations of the microbeam and the air film, and real eigenvalues describing the relaxation of the pressure field. The physical content of the complex modes is analyzed through their phase, and the corresponding distribution of the dissipation over the film. The reduced model is then compared to the full discretized system for clamped--clamped and clamped--free microbeams at selected pressures.
	
	The paper is organized as follows. Section \ref{sec:formulation} presents the governing equations. Section \ref{sec:evp} describes the discretization, the static solution and the coupled eigenvalue problem. Section \ref{sec:modes} analyses the complex modes, validates the frequencies and quality factors against experimental and numerical results from the literature, and examines their phase and dissipation content. Section \ref{sec:rom} develops the reduced-order model, and Section \ref{sec:dynamics} assesses its convergence and accuracy in the forced nonlinear regime. Conclusions are drawn in Section \ref{sec:conclusions}.
	
	%%%%%%%%%%%%%%%%%%%%%%%%%%%%%%%%%%%%%%%%%%%%%%%%%%%%%%%%%%%%%%%%%%%%%%%%%%%%%%%%%%%%%%%%%%%%%%%%%%%%%%%%%%%%%%%%%%%%%%%%%%%%%%%%%
	\section{Problem formulation and governing equations}\label{sec:formulation}
	This section presents the two microbeam configurations studied in this work and the equations governing their coupled dynamics. The structural and squeeze-film models are introduced in nondimensional form with the related assumptions. 
	
	\begin{figure}[H]
		\centering
		\includegraphics[width=0.7\columnwidth]{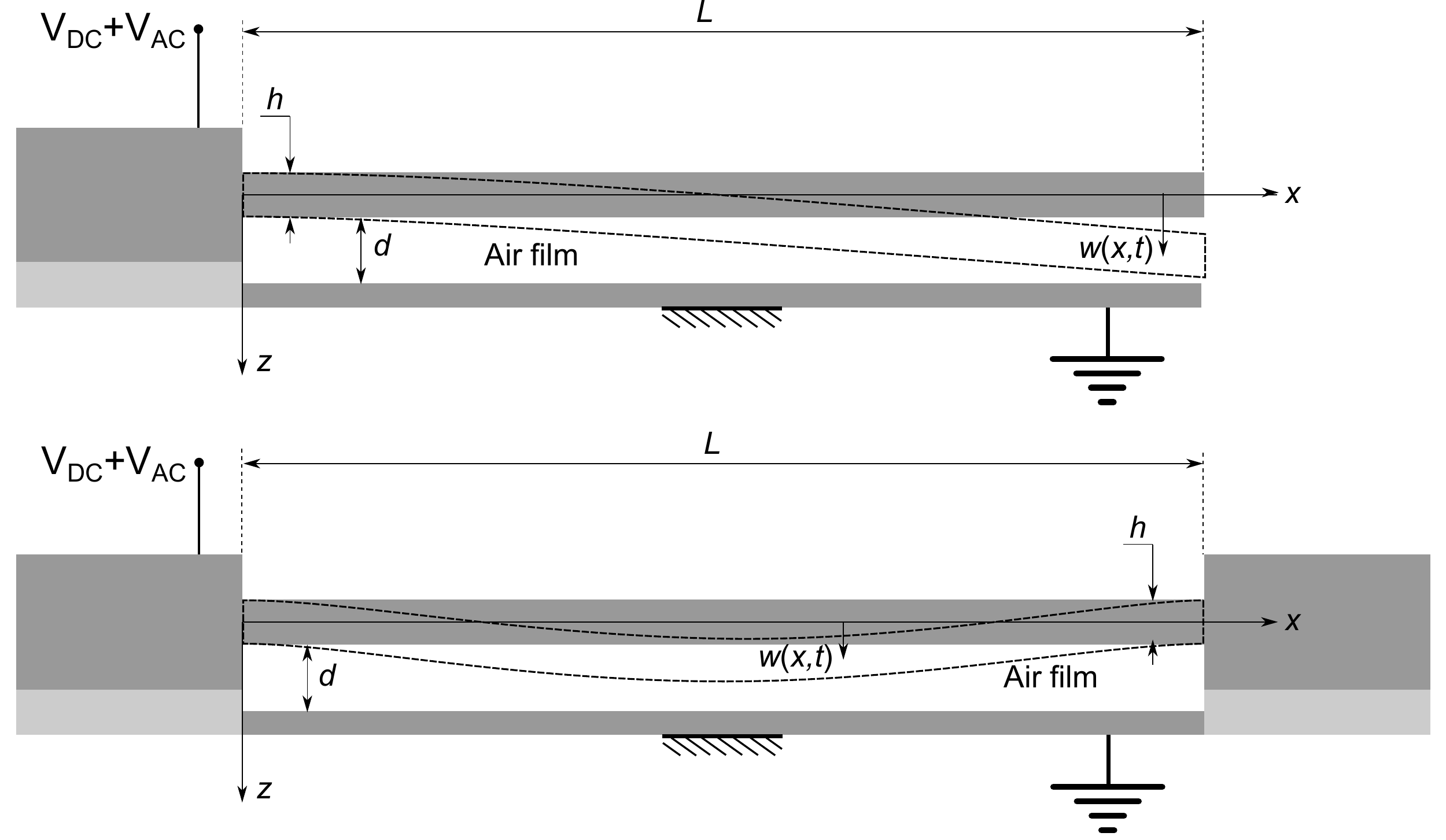}
		\caption{Schematics of the electrostatically actuated microbeams separated from the substrate electrode by an air film: clamped-free (top) and clamped-clamped (bottom).} 
		\label{fig1}
	\end{figure}

	Two configurations are considered in this work, the clamped--clamped (CC) and the clamped--free (CF) boundary conditions. Both are shown in Figure \ref{fig1} along with the substrate electrode and the applied voltages. $L$, $b$ and $h$ are the length, width and thickness of the microbeam, and $d$ is the initial gap.

	The transverse motion of each microbeam is described by a nonlinear equation that accounts for moderately large deflections, residual axial stresses and the nonlinear dependence of the electrostatic force on the gap, including the fringing-field correction. The gas film trapped between the microbeam and the substrate electrode is described by the nonlinear two-dimensional Reynolds equation \citep{hamrock2004fundamentals}, which governs the pressure distribution over the whole film and couples it to the motion of the microbeam.
	
	The microbeam is modeled as an Euler--Bernoulli beam and the elastic modulus listed in Table \ref{tab1} is used as an effective value. The mid-plane stretching induced by the clamped--clamped boundary conditions is introduced through von K\'arm\'an strain, it is absent for the CF microbeam. The gas film is described by the isothermal compressible Reynolds equation, which presumes a film thickness small compared with the lateral dimensions ($d \ll L,\,b$), laminar flow, and negligible gas inertia. Rarefaction is accounted for through an effective viscosity. The electrostatic force is evaluated in the parallel-plate approximation corrected for fringing fields.
	
	The nondimensional transverse deflection $w(x,t)$ and nondimensional relative pressure $p(x,y,t)$ for the CC and CF microbeams are supposed to be governed by the following equations and associated boundary conditions \cite{sassi2018square,sassi2018strong}:
	
	\textit{\underline{CC microbeam:}}
	\begin{align} 
		&\ddot w + \gamma \dot w + w_{,xxxx} = w_{,xx}  \left( N - \alpha _1 \int_0^1 w \, w_{,xx} \, dx \right)  \label{eq1} \\
		& \hspace{2.8cm}  + {\alpha _2}\frac{{{{\left( {{V_{DC}} + V_{AC}(t)} \right)}^2}}}{{{{\left( {1 - w} \right)}^2}}}\left( {1 + \alpha _3 (1 - w)} \right) -{\alpha _4}\int_0^1 {p \, dy} \nonumber \\
		& \left[ {{(1-w)^3} (1+p)  p_{,x}} \right]_{,x} + \alpha_5  \left[ {{(1-w)^3} (1+p) p_{,y}} \right]_{,y} = \sigma \Big((1-w) \dot{p} - (1+p)\dot{w} \Big) \label{eq2}
	\end{align}
	subjected to the nondimensional boundary conditions
	\begin{align}
		&w(0,t) =0, \; w(1,t) = 0, \; w_{,x} (0,t) =0 \mbox{ and }  w_{,x}(1,t) = 0 \label{eq3} \\
		&p(x,0,t) =0,\; p(x,1,t) =0,\;  p_{,x}(0,y,t) =0 \mbox{ and } p_{,x}(1,y,t) = 0 \label{eq4}
	\end{align}
	
	\textit{\underline{CF microbeam:}}
	\begin{align} 
		& \ddot w + \gamma\dot w + w_{,xxxx} = \frac{{{\alpha _2}{{\left( {{V_{DC}} + V_{AC}(t)} \right)}^2}}}{{{{\left( {1 - w} \right)}^2}}} \left( {1 + \alpha _3 (1 - w)} \right) -{\alpha _4}\int_0^1 {p \, dy} \label{eq5}\\
		& \left[ {{(1-w)^3} (1+p)  p_{,x}} \right]_{,x} + \alpha_5  \left[ {{(1-w)^3} (1+p) p_{,y}} \right]_{,y} = \sigma \Big((1-w) \dot{p} - (1+p)\dot{w} \Big) \label{eq6}
	\end{align}  
	subjected to the nondimensional boundary conditions
	\begin{align}
		&w(0,t) =0,\; w_{,x}(0,t) = 0,\; w_{,xx}(1,t) =0  \mbox{ and } w_{,xxx}(1,t) = 0 \label{eq7} \\
		&p(x,0,t) =0,\; p(x,1,t) =0, \; p(1,y,t) =0 \mbox{ and } p_{,x}(0,y,t) = 0 \label{eq8}
	\end{align}
	where the over-dots denote partial derivatives with respect to the nondimensional time, the subscripts $x$ and $y$ after a comma denote spatial partial derivatives, and $V_{AC}(t) = v_{ac}\,\cos(\Omega t)$ is the harmonic component of the applied voltage, $\Omega$ being its nondimensional excitation frequency. The mid-plane stretching term of Equation \eqref{eq1} is written in the form $-\int_0^1 w\,w_{,xx}\,dx$, obtained from $\int_0^1 w_{,x}^2\,dx$ by integration by parts, since the deflection vanishes at both clamped ends. This form avoids squaring the discretized slope and is used throughout the paper.
	
	In Equations \eqref{eq1} to \eqref{eq8}, $\gamma$ denotes the nondimensional coefficient of structural damping (material and anchor), assumed viscous and uniform along the microbeam. The effect of the gas film, dissipative and elastic, affects the structural equations through the coupling term $\alpha_4 \int_0^1 p\,dy$. The following nondimensional variables and parameters are used:
	\begin{equation}  \label{eq9}
		\begin{aligned}
			&x = \frac{{\hat x}}{L},\,\,y = \frac{{\hat y}}{b},\,\,w = \frac{{\hat w}}{d},\,\, t = \frac{{\hat t}}{\tau} , \,\, p= \frac{{\hat p-p_a}}{p_a}, \,\, \tau= \sqrt{\frac{{12 \rho {L^4}}}{{Eh^2}}},  \,\, N = \frac{{\hat N{L^2}}}{{EI}}, \,\, \gamma = \frac{\hat c L^4}{EI\,\tau} \\
			&{\alpha _1} = 6\frac{{{d^2}}}{{{h^2}}}, \,\, {\alpha _2} = \frac{{6\varepsilon {L^4}}}{{E{h^3}{d^3}}}, \,\,  {\alpha _3} = \frac{{0.65}}{{b/d}}, \,\, {\alpha _4} =\frac{12  p_a L^4 }{d E h^3} , \,\, {\alpha _5} = \frac{L^2}{b^2} , \,\,  \sigma=\frac{12 \mu_{eff} L^2}{\tau p_a d^2}
		\end{aligned}
	\end{equation}
	where hats denote dimensional quantities, $L$, $b$ and $h$ are the length, width and thickness of the microbeam, $d$ is the initial gap, $E$ and $\rho$ are the Young's modulus and density of the microbeam, $I = bh^3/12$ is the second moment of area of its cross-section, $\varepsilon$ is the permittivity of the gap medium and $\hat c$ is the structural damping coefficient per unit length. The relative pressure $p$ is defined with respect to the ambient pressure $p_a$, $\hat p$ being the absolute pressure in the gap, and $\hat N$ is the resultant axial force arising from residual stresses. The effective viscosity $\mu_{eff}$ accounts for gas rarefaction and is expressed as \citep{Veijola1995}:
	\begin{equation} \label{eq10}
		\mu _{eff}= \frac{\mu}{1+9.638 \, {K_n}^{1.159}}
	\end{equation}
	where $\mu$ is the dynamic viscosity of air at standard pressure and temperature, whose value is given in Table \ref{tab1}, and $K_n$ is the Knudsen number, given by
	\begin{equation} \label{eq11}
		K_n=\frac{\lambda_0 \, p_0 }{p_a d}
	\end{equation}
	here $\lambda_0=65 \times 10^{-9}$m is the mean free path and $p_0=1.01325 \times 10^5$Pa is the standard pressure.
	
	As expected the coupled systems \eqref{eq1}--\eqref{eq2} and \eqref{eq5}--\eqref{eq6} shows asymmetric coupling between the deflection and the pressure fields. The microbeam is loaded by the pressure $p(x,y,t)$ through the term $\alpha_4 \int_0^1 p\,dy$ in Equations \eqref{eq1} and \eqref{eq5}, whereas the film is driven by the velocity of the microbeam through the term $\sigma(1+p)\dot w$ in Equations \eqref{eq2} and \eqref{eq6}. The linearized operator associated with this system is therefore non-symmetric and non-self-adjoint, so that its eigenvalues are in general complex and its left and right eigenvectors are distinct. Physically, the deflection and the pressure do not reach their extrema simultaneously so that each coupled mode carries an intrinsic phase lag between the structural and fluid fields, in addition to its own decay rate. Capturing this behavior is what motivates the complex-mode basis developed in the following sections.
	
	Since the squeeze number $\sigma$ is defined with respect to the structural timescale $\tau$, the squeeze number relevant to a motion of nondimensional frequency $\omega$ is $\sigma\omega$. A low value corresponds to a film that relaxes rapidly compared with the microbeam motion, whereas a high value indicates a trapped film whose compression cannot be relieved within a period.
	
	%%%%%%%%%%%%%%%%%%%%%%%%%%%%%%%%%%%%%%%%%%%%%%%%%%%%%%%%%%%%%%%%%%%%%%%%%%%%%%%%%%%%%%%%%%%%%%%%%%%%%%%%%%%%%%%%%%%%%%%%%%%%%%%%%
	\section{Discretization and coupled eigenvalue problem}\label{sec:evp}
	This section describes the spatial discretization of the coupled problem, the static equilibrium about which the equations are linearized, and the resulting eigenvalue problem.
	
	\subsection{Discretization using the DQM and LaDQM}
	The spatial discretization is carried out using the differential quadrature method (DQM), in which the derivative of a function at a grid point is approximated by a weighted linear sum of the function values at all points along the corresponding coordinate direction \citep{bellman1971differential}.  The method delivers accurate results with a relatively small number of grid points \citep{najar2005modeling}. This property is important especially for the two-dimensional fluid domain.
	
	A well-documented difficulty of the classical DQM arises for fourth-order problems such as the Euler--Bernoulli beam, where two boundary conditions must be enforced at each end while only one discrete equation is available per grid point \citep{wang2004new}. Here we adopt the locally adaptive differential quadrature method (LaDQM) \citep{ng2009accuracy}, in which fictitious grid points located outside the physical domain carry the additional degrees of freedom required to impose the boundary conditions exactly. Beyond accuracy, this choice also reduces the creation of spurious complex eigenvalues present in the classical DQM formulation \citep{ng2009accuracy}.
	
	Accordingly, the structural equations \eqref{eq1} and \eqref{eq5}, together with their boundary conditions \eqref{eq3} and \eqref{eq7}, are discretized using LaDQM, with one fictitious point added beyond each end of the microbeam. Because the boundary conditions are carried by the fictitious points, no interior grid point has to be sacrificed to enforce them, and every structural node remains available for coupling with the fluid domain through the velocity of the microbeam at each grid point along $x$.
	
	The Reynolds equations \eqref{eq2} and \eqref{eq6}, being of second order in both spatial directions, admit a single boundary condition per edge and are therefore discretized with the classical DQM on a two-dimensional grid. All spatial derivatives in Equations \eqref{eq1} to \eqref{eq8} are replaced by weighted sums, using the LaDQM weighting coefficients $A^{(r)}_{ij}$ for the $r^{th}$-order derivative in the structural domain and the corresponding classical DQM coefficients $B^{(r)}_{ij}$ in the fluid domain (see \ref{ApI}). Both domains share the same Chebyshev--Gauss--Lobatto (CGL) grid along $x$, which avoids any interpolation in the coupling terms. The grid points are given by
	\begin{equation*}
		x_i=\frac12\left( 1-\cos \frac{i-1}{n-1}\pi\right) \mbox{ and }
		y_i=\frac12\left( 1-\cos \frac{i-1}{n-1}\pi\right)  \mbox{ ,  } i=1,\ldots,n
	\end{equation*}
	where $n$ is the number of grid points in the $x$ and $y$ directions.
	
	The integral terms in Equations \eqref{eq1} and \eqref{eq5} are evaluated on the same grid, using the quadrature interpolation obtained by integrating the same DQM Lagrange polynomials \citep{bert1996differential}. The corresponding weights $T_i$ are given in Equation \eqref{eqNC} in \ref{ApI}. These weights are used only for the integrals that belong to the physical model, namely the mid-plane stretching term and the transverse integral of the pressure.
	
	Symmetry of the solution is only applied in the $y$-direction, where both film edges are open, and the pressure field is symmetric about $y=1/2$ for any deflection shape. No symmetry is imposed along $x$. We end up with the following set of nonlinear ODEs in time and their associated algebraic boundary conditions:
	
	\textit{\underline{CC microbeam:}}
	\begin{align} 
		&\ddot w_i + \gamma\dot w_i + \sum_{j=0}^{n+1}{A_{ij}^{(4)} w_j} = \left( \sum_{j=0}^{n+1}{A_{ij}^{(2)} w_j} \right)   \left[ {{N} - {\alpha _1}  \sum_{j=1}^{n}{T_j w_j \left( \sum_{k=0}^{n+1}{A_{jk}^{(2)} w_k}\right) }  } \right] \label{eq12} \\ 
		& \hspace{1.3cm} + {\alpha _2}\frac{{{{\left( {{V_{DC}} + V_{AC}(t)} \right)}^2}}}{{{{\left( {1 - w_i} \right)}^2}}}\left( {1 + \alpha_3 (1 - w_i)} \right)   - {\alpha _4} \sum_{j=1}^{n}{T_j p_{ij} }  \nonumber \\
		& \hspace{1.3cm} i=2 \cdots n-1  \nonumber \\
		& (1-w_i)^3 \left(  \sum_{k=1}^{n}{B_{ik}^{(1)} p_{kj}}\right)^2  -3 (1-w_i)^2 (1+p_{ij}) \left( \sum_{k=0}^{n+1}{A_{ik}^{(1)} w_k}\right) \left(  \sum_{k=1}^{n}{B_{ik}^{(1)} p_{kj}}\right) \label{eq13} \\
		& \hspace{1.3cm} + (1-w_i)^3 (1+p_{ij}) \left(  \sum_{k=1}^{n}{B_{ik}^{(2)} p_{kj}}\right)  + \alpha_5  (1-w_i)^3 \left(  \sum_{k=1}^{n}{B_{jk}^{(1)} p_{ik}}\right)^2 \nonumber \\
		& \hspace{1.3cm} + \alpha_5 (1-w_i)^3 (1+p_{ij})  \left(  \sum_{k=1}^{n}{B_{jk}^{(2)} p_{ik}}\right) = \sigma (1-w_i) \dot p_{ij} - \sigma (1+p_{ij}) \dot{w_i} \nonumber\\
		& \hspace{1.3cm} i=2 \cdots n-1 \; \mbox{ ; }\; j=2 \cdots \frac{n+1}{2} \nonumber 
	\end{align}
	subjected to the nondimensional boundary conditions
	\begin{align}
		&w_1 =0,\;  w_n = 0,\; \sum_{j=0}^{n+1}{A_{1j}^{(1)} w_j} =0  \mbox{ and } \sum_{j=0}^{n+1}{A_{nj}^{(1)} w_j} = 0 \label{eq14} \\
		&p_{i1} =0  \mbox{ and } p_{in} =0 \; \mbox{ while } i=1 \cdots n \label{eq15} \\
		& \sum_{k=1}^{n}{B_{1k}^{(1)} p_{kj}} =0 \mbox{ and } \sum_{k=1}^{n}{B_{nk}^{(1)} p_{kj}} = 0 \; \mbox{ while } j=1 \cdots n \nonumber
	\end{align}
	
	\textit{\underline{CF microbeam:}}
	\begin{align} 
		&\ddot w_i + \gamma\dot w_i + \sum_{j=0}^{n+1}{A_{ij}^{(4)} w_j} = {\alpha _2}\frac{{{{\left( {{V_{DC}} + V_{AC}(t)} \right)}^2}}}{{{{\left( {1 - w_i} \right)}^2}}}\left( {1 + \alpha_3 (1 - w_i)} \right) - {\alpha _4} \sum_{j=1}^{n}{T_j p_{ij} }  \label{eq16} \\
		& \hspace{1.3cm} i=2 \cdots n-1  \nonumber \\
		& (1-w_i)^3 \left(  \sum_{k=1}^{n}{B_{ik}^{(1)} p_{kj}}\right)^2  -3 (1-w_i)^2 (1+p_{ij}) \left( \sum_{k=0}^{n+1}{A_{ik}^{(1)} w_k}\right) \left(  \sum_{k=1}^{n}{B_{ik}^{(1)} p_{kj}}\right) \label{eq17} \\
		& \hspace{1.3cm} + (1-w_i)^3 (1+p_{ij}) \left(  \sum_{k=1}^{n}{B_{ik}^{(2)} p_{kj}}\right)  + \alpha_5  (1-w_i)^3 \left(  \sum_{k=1}^{n}{B_{jk}^{(1)} p_{ik}}\right)^2 \nonumber \\
		& \hspace{1.3cm} + \alpha_5 (1-w_i)^3 (1+p_{ij})  \left(  \sum_{k=1}^{n}{B_{jk}^{(2)} p_{ik}}\right) = \sigma (1-w_i) \dot p_{ij} - \sigma (1+p_{ij}) \dot{w_i} \nonumber \\
		& \hspace{1.3cm} i=2 \cdots n-1 \; \mbox{ ; }\; j=2 \cdots \frac{n+1}{2} \nonumber 
	\end{align}
	subjected to the nondimensional boundary conditions
	\begin{align}
		&w_1 =0,\;  \sum_{j=0}^{n+1}{A_{1j}^{(1)} w_j} =0,\; \sum_{j=0}^{n+1}{A_{nj}^{(2)} w_j} =0  \mbox{ and } \sum_{j=0}^{n+1}{A_{nj}^{(3)} w_j} = 0 \label{eq18} \\
		&p_{i1} =0  \mbox{ and } p_{in} =0 \; \mbox{ while } i=1 \cdots n \label{eq19} \\
		& \sum_{k=1}^{n}{B_{1k}^{(1)} p_{kj}} =0 \mbox{ and } p_{nj} = 0 \; \mbox{ while } j=1 \cdots n \nonumber 
	\end{align}
	where $w_i=w(x_i,t)$ and $p_{ij}=p(x_i,y_j,t)$.
	
	Since all the boundary conditions in Equations \eqref{eq14}, \eqref{eq15}, \eqref{eq18} and \eqref{eq19} are linear, they are solved directly to express the boundary and fictitious nodes in terms of the interior ones, and are therefore no longer written explicitly in what follows.
	
	\subsection{Static response}\label{Subsec_Static}
	The static response, denoted by $u(x)$, satisfies the equations obtained from Equations \eqref{eq1}, \eqref{eq3}, \eqref{eq5} and \eqref{eq7} by suppressing the time derivatives and the AC component of the voltage. At equilibrium the relative pressure is zero, and the film has no effect on the static problem. The governing equations are, for the two configurations,
	
	\textit{\underline{CC microbeam:}}
	\begin{align} 
		&u_{,xxxx} = u_{,xx}  \left( N - \alpha _1 \int_0^1 u \, u_{,xx} \, dx \right) + {\alpha _2}\frac{{V_{DC}}^2}{{{{\left( {1 - u} \right)}^2}}}\left( {1 + \alpha _3 (1 - u)} \right) \label{eq20} \\
		&u(0) =0, \; u(1) = 0, \; u_{,x} (0) =0 \mbox{ and }  u_{,x}(1) = 0 \label{eq21}
	\end{align}
	
	\textit{\underline{CF microbeam:}}
	\begin{align} 
		& u_{,xxxx} = \frac{{{\alpha _2}{{{{V_{DC}}}}^2}}}{{{{\left( {1 - u} \right)}^2}}} \left( {1 + \alpha _3 (1 - u)} \right) \label{eq22}\\
		& u(0) =0,\; u_{,x}(0) = 0,\; u_{,xx}(1) =0  \mbox{ and } u_{,xxx}(1) = 0 \label{eq23}
	\end{align}
	
	Applying the LaDQM to these equations, we end up with
	
	\textit{\underline{CC microbeam:}}
	\begin{align} 
		&\sum_{j=0}^{n+1}{A_{ij}^{(4)} u_j} = \left( \sum_{j=0}^{n+1}{A_{ij}^{(2)} u_j} \right)   \left[ {{N} - {\alpha _1}  \sum_{j=1}^{n}{T_j u_j \left( \sum_{k=0}^{n+1}{A_{jk}^{(2)} u_k}\right) }  } \right]  \label{eq24} \\  
		& \hspace{1.9cm} + {\alpha _2}\frac{{{{{{V_{DC}}}}^2}}}{{{{\left( {1 - u_i} \right)}^2}}}\left( {1 + \alpha_3 (1 - u_i)} \right)   \hspace{1.3cm} i= 2 \cdots n-1  \nonumber \\
		& u_1 =0,\;  u_n = 0,\; \sum_{j=0}^{n+1}{A_{1j}^{(1)} u_j} =0  \mbox{ and } \sum_{j=0}^{n+1}{A_{nj}^{(1)} u_j} = 0 \label{eq25}
	\end{align}
	
	\textit{\underline{CF microbeam:}}
	\begin{align} 
		&\sum_{j=0}^{n+1}{A_{ij}^{(4)} u_j} = {\alpha _2}\frac{{{{ {{V_{DC}}}}^2}}}{{{{\left( {1 - u_i} \right)}^2}}}\left( {1 + \alpha_3 (1 - u_i)} \right)  \hspace{1.3cm} i=2 \cdots n-1 \label{eq26}  \\
		& u_1 =0,\;  \sum_{j=0}^{n+1}{A_{1j}^{(1)} u_j} =0,\; \sum_{j=0}^{n+1}{A_{nj}^{(2)} u_j} =0  \mbox{ and } \sum_{j=0}^{n+1}{A_{nj}^{(3)} u_j} = 0 \label{eq27}
	\end{align}
	where $u_i=u(x_i)$ is the nodal static response.
	
	\begin{table}[H]
		\centering
		\caption{Microbeam parameters \citep{tilmans1994electrostatically,nayfeh2004squeeze,pandey2007effect,chaterjee2012squeeze}. The viscosity of air is taken, for each microbeam, as in the reference works used for validation.}
		\label{tab1}
		\begin{tabular}{ccccccccc}
			\hline
			Beam & $L$ & $b$ & $h$ & $d$ & $E$ & $\rho$ & $\hat N$ & $\mu$ \\ 
			& ($\mu$m) & ($\mu$m) & ($\mu$m) & ($\mu$m) & (GPa) & (kg/m$^{3}$) & (mN) & (Ns/m$^2$) \\ \hline
			CC  &     310      &     100      &     1.5      &     1.18    &    166    &        2332         &      0.9      & $1.983\times10^{-5}$\\
			CF  &     350      &      22      &      4       &     1.4      &    160    &        2330         &       0       & $1.8\times10^{-5}$ \\ \hline
		\end{tabular}
	\end{table}
	
\begin{figure}[H]
		\centering
		\includegraphics[width=0.6\columnwidth]{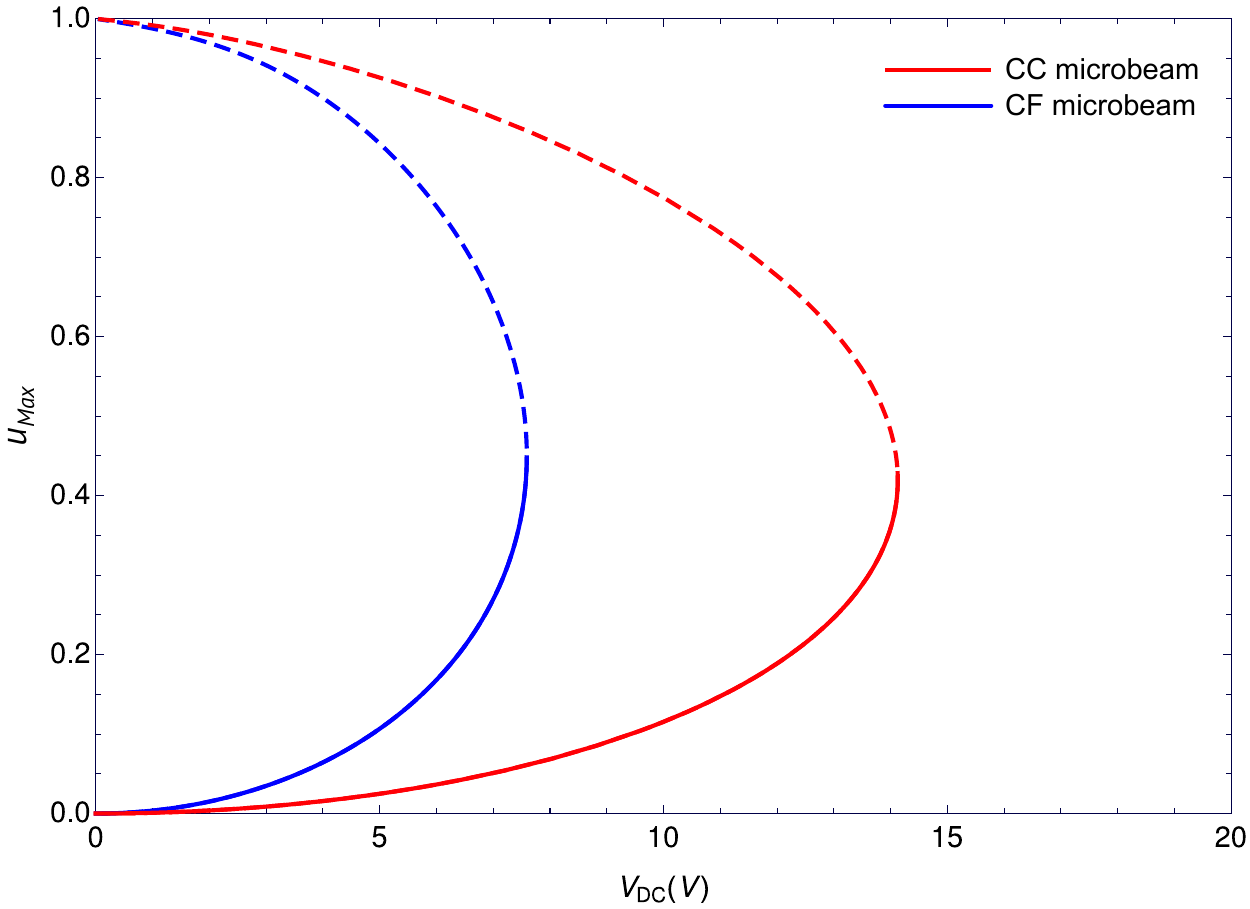}
		\caption{Maximum static deflection of the CC and CF microbeams of Table \ref{tab1}, as a function of the applied DC voltage. Solid lines denote the stable branch and dashed lines the unstable one, the two meeting at the static pull-in voltage.} 
		\label{fig2}
	\end{figure}

	The nodal static deflections $u_i$ are obtained by solving the nonlinear algebraic equations \eqref{eq24}--\eqref{eq25} for the CC microbeam and \eqref{eq26}--\eqref{eq27} for the CF microbeam, using a Newton--Raphson procedure following \citet{najar2005modeling}. The resulting solution serves as the operating point about which the eigenvalue problem is linearized. For the parameters of Table \ref{tab1}, the maximum static deflection $u_{Max}$ is shown in Figure \ref{fig2} as a function of the DC voltage. Two branches of static solutions are obtained, the upper one being unstable, and they meet at the static pull-in voltage. In the dynamic analysis of Section \ref{sec:dynamics}, the static solution is computed at the root-mean-square voltage $V_{RMS}=\sqrt{V_{DC}^2+\frac12 v_{ac}^2}$ rather than at $V_{DC}$.
	
	\subsection{Linearized coupled eigenvalue problem}
	The linearized coupled eigenvalue problem associated with Equations \eqref{eq1}--\eqref{eq4} for the CC microbeam and \eqref{eq5}--\eqref{eq8} for the CF microbeam is obtained by linearizing the equations about the static deflection $u$ and the relative pressure $p=0$. The deflection is written as $w(x,t)=u(x)+q(x,t)$, where $q$ is the deflection dynamic component about the static position.
	
	The nonlinear terms are expanded to first order in $q$ and $p$, and the terms that satisfy the static equations \eqref{eq20} and \eqref{eq22} are eliminated, which yields
	
	\textit{\underline{CC microbeam:}}
	\begin{align} 
		&\ddot q + \gamma\dot q + q_{,xxxx} = - \alpha _1 u_{,xx}   \int_0^1 \left( u q_{,xx} + q u_{,xx} \right)  dx  + q_{,xx}  \left( N - \alpha _1 \int_0^1 u u_{,xx} dx \right)  \label{eq28} \\
		& \hspace{2.8cm}  + {\alpha _2}\frac{{{{{{V_{DC}}}}^2}}}{{{{\left( {1 - u} \right)}^3}}}\left( {2 + \alpha _3 (1 - u)} \right)q -{\alpha _4}\int_0^1 {p \, dy} \nonumber \\
		& \left[ {{(1-u)^3}  p_{,x}} \right]_{,x} + \alpha_5  {{(1-u)^3} p_{,yy}} = \sigma \Big((1-u) \dot{p} - \dot{q} \Big) \label{eq29}
	\end{align}
	with boundary conditions
	\begin{align}
		&q(0,t) =0, \; q(1,t) = 0, \; q_{,x} (0,t) =0 \mbox{ and }  q_{,x}(1,t) = 0 \label{eq30} \\
		&p(x,0,t) =0,\; p(x,1,t) =0,\;  p_{,x}(0,y,t) =0 \mbox{ and } p_{,x}(1,y,t) = 0 \label{eq31}
	\end{align}
	
	\textit{\underline{CF microbeam:}}
	\begin{align} 
		& \ddot q + \gamma \dot q + q_{,xxxx} = {\alpha _2}\frac{{{{{{V_{DC}}}}^2}}}{{{{\left( {1 - u} \right)}^3}}}\left( {2 + \alpha _3 (1 - u)} \right)q -{\alpha _4}\int_0^1 {p \, dy} \label{eq32}\\
		& \left[ {{(1-u)^3}  p_{,x}} \right]_{,x} + \alpha_5  {{(1-u)^3} p_{,yy}} = \sigma \Big((1-u) \dot{p} - \dot{q} \Big) \label{eq33}
	\end{align}  
	with boundary conditions
	\begin{align}
		&q(0,t) =0,\; q_{,x}(0,t) = 0,\; q_{,xx}(1,t) =0  \mbox{ and } q_{,xxx}(1,t) = 0 \label{eq34} \\
		&p(x,0,t) =0,\; p(x,1,t) =0, \; p(1,y,t) =0 \mbox{ and } p_{,x}(0,y,t) = 0 \label{eq35}
	\end{align}
	
	Next we apply the LaDQM to discretize the linearized dynamic problem given by Equations \eqref{eq28}-\eqref{eq35}. The resulting linear algebraic equations are given for each case in \ref{ApIevp}. Collecting the variable and applying the associated boundary and symmetry conditions, the system can be written in matrix form as follows:
	\begin{equation} \label{eq44}
		\bf M \ddot x +  C \dot x + K(u) x =  0, \; {\bf x}= \left( \begin{array}{c} {\bf{q}}  \\ \bf p  \end{array}   \right)
	\end{equation}
	where $\bf M$, $\bf C$ and $\bf K(u)$ are $\left( (n_w+n_p)\times(n_w+n_p)\right) $ matrices corresponding to mass, damping and stiffness, respectively, $\bf u$ is the static solution vector, and $ {\bf x}$ is the unknown displacement and pressure vector with dimension $\left( (n_w + n_p) \times 1\right)$. The dimensions of the different matrices are $n_w = n-2$ and $n_p = (n-2)\frac{n-1}{2}$ for both the CC and the CF cases.
	
	The associated eigenvalue problem is obtained by letting $q_i=Q_i\,e^{\lambda t}$ and $p_{ij}(t)=P_{ij}\,e^{\lambda t}$, where $Q_i=Q(x_i)$ and $P_{ij}=P(x_i,y_j)$ are node functions at the DQM grid points. Therefore, the unknown vector can be expressed as ${\bf x = X } \,e^{\lambda t} $ and Equation \eqref{eq44} is transformed to
	\begin{equation} \label{eq45}
		\left[  \lambda^2 {\bf M} + \lambda {\bf C+ K(u) }\right] {\bf X} = \bf 0, \; {\bf X}= \left( \begin{array}{c} \bf Q \\ \bf P  \end{array}   \right)
	\end{equation}
	where the eigenvector $ {\bf X}$ is composed by two partitions: a vector $\bf Q$ with dimension $\left( n_w \times 1\right)$ followed by a vector $\bf P$ with dimension $\left( n_p \times 1\right)$ obtained by reshaping the $P_{ij}$ components, line by line, in a vector form. More specifically:
	$$ {\bf X}= \left( Q_2,Q_3 \cdots Q_{n-1},P_{22},P_{23} \cdots P_{n-1;\frac{n+1}{2}} \right)^T \mbox{ with dimension } \left( (n-2)\frac{n+1}2 \times 1 \right) $$
	
	The mass matrix $\bf M$ in Equation \eqref{eq44} is singular, it contains an identity block associated with the structural degrees of freedom, all its other entries being zero, since the Reynolds equation is of first order in time. The damping and stiffness matrices are both non-symmetric. Part of this asymmetry is a property of the collocation itself, the DQM weighting matrices being non-symmetric even for the uncoupled structure. The asymmetry that is physical, and that generates the complex modes, is coming from the coupling blocks, discussed below. The pressure loads the structure through the stiffness matrix, whereas the structure drives the pressure through the damping matrix.
	
	To solve for the eigenvalues and eigenvectors of Equation \eqref{eq45}, a generalized eigenvalue problem should be written in the state space. Following \cite{meirovitch2010fundamentals}, we introduce the state vector ${\bf y}$ and transform Equation \eqref{eq44} to 
	\begin{equation} \label{eq46}
		{\bf B y }=  {\bf A \dot y} , \; {\bf y} = \left( \begin{array}{c} \bf x \\ \bf \dot x  \end{array} \right)
	\end{equation}
	where
	$$\bf B = 
	\left[ \begin{array}{cc} 
		\bf 0 & \bf I \\
		\bf -K & \bf -C 
	\end{array} \right]
	\mbox{  and  } 
	\bf A = \left[ \begin{array}{cc} 
		\bf I & \bf 0 \\
		\bf 0 & \bf M 
	\end{array} \right] $$
	
	Assuming that ${\bf y} = {\bf Y } e^{\lambda t}$, Equation \eqref{eq46} is transformed into a generalized eigenvalue problem associated to Equation \eqref{eq44}, it can be written as 
	\begin{equation} \label{eq47}
		{\bf B Y }= \lambda {\bf A Y} 
	\end{equation}
	
	The eigenvalues of \eqref{eq47} and \eqref{eq45} are the same \citep{yang1995modal,omenzetter2012interaction}, and their corresponding eigenfunction can be linked using the following expression:
	\begin{equation} \label{eq48}
		{\bf Y}= \left( \begin{array}{c} {\bf X} \\ \lambda {\bf X } \end{array}  \right) = \left( \begin{array}{c} {\bf Q} \\ {\bf P} \\ \lambda {\bf Q } \\ \lambda {\bf P } \end{array}  \right)
	\end{equation}
	
	The real matrix $\bf B$ is non-symmetric and the matrix $\bf A$ is singular, the diagonal entries of $\bf M$ associated with the pressure degrees of freedom being zero. The eigenvalue problem \eqref{eq47} is therefore transformed to remove this singularity:
	\begin{equation} \label{eq49}
		\left[ \begin{array}{cc} 
			{\bf B}_{11}^{((2n_w+n_p) \times (2n_w+n_p))} & {\bf B}_{12}^{((2n_w+n_p) \times n_p)} \\
			{\bf B}_{21}^{(n_p \times (2n_w+n_p))} & {\bf B}_{22}^{(n_p \times n_p)} 
		\end{array} \right] \left( \begin{array}{c} 
			\bf \Phi  \\
			\lambda {\bf P }
		\end{array} \right)=  \lambda  \left[ \begin{array}{cc} 
			{\bf A}_{11} & \bf 0 \\
			\bf 0 & \bf \bf 0 
		\end{array} \right] \left( \begin{array}{c} 
			\bf \Phi  \\
			\lambda {\bf P }
		\end{array} \right)
	\end{equation}
	where ${\bf \Phi} = \left( \begin{array}{c} {\bf \Phi^1}^{(n_w \times 1)} \\ {\bf \Phi^2}^{(n_p \times 1)} \\ {\bf \Phi^3}^{(n_w \times 1)} \end{array}  \right) = \left( \begin{array}{c} {\bf Q} \\ {\bf P} \\ \lambda {\bf Q } \end{array}  \right)$.
	
	In addition, it can be shown that ${\bf A}_{11}={\bf I}^{(2n_w+n_p \times 2n_w+n_p)}$, because the mass matrix $\bf M$ contains only 1 or 0 at the diagonal. Consequently, the generalized eigenvalue problem \eqref{eq47} is transformed to a classical one, given by the following equations:
	\begin{align} 
		&{\bf B}_{11} {\bf \Phi} + \lambda {\bf B}_{12} {\bf P } = \lambda {\bf \Phi} \label{eq50} \\
		&{\bf B}_{21} {\bf \Phi} + \lambda {\bf B}_{22} {\bf P } = {\bf 0} ^{(n_p \times 1)}  \nonumber
	\end{align}
	That is,
	\begin{equation} 
		{\bf \hat B \Phi } = \lambda {\bf \Phi} \label{eq51}
	\end{equation}
	where $\bf \hat B = \left[  {\bf B}_{11} - {\bf B}_{12} {\bf B}_{22}^{-1} {\bf B}_{21} \right] $ is a $\left( (2n_w+n_p)\times(2n_w+n_p)\right) $ matrix.
	
	The eigenvalue problem \eqref{eq51} is equivalent to the original problem \eqref{eq45}. The condensation requires ${\bf B}_{22}$ to be invertible, which is guaranteed here, since ${\bf B}_{22}=-{\bf C}_{22}$ and ${\bf C}_{22}$ is a diagonal matrix with elements proportional to $\sigma(1-u_i)$, which do not vanish below pull-in.
	
	However, the matrix associated with the new eigenvalue problem in \eqref{eq51} is real but non-symmetric, and consequently non-self-adjoint. Its right eigenvectors ${\bf \Phi}$ are therefore not mutually orthogonal, and in general complex. Because ${\bf \hat B}$ is real, the eigenvalues are either real or occur in complex-conjugate pairs. These two sets have distinct physical meanings. Real eigenvalues correspond to overdamped, non-oscillatory motions in which the pressure field relaxes monotonically, they are a direct product of the compressibility and viscosity of the air film and have no counterpart in the undamped structural problem, the absence of oscillation reflects the absence of inertia in the Reynolds model. Complex-conjugate pairs of the eigenvalues correspond to underdamped coupled motions, in which the real part describes the decay rate and the imaginary part the damped frequency \citep{chen2020generalized}. The complex eigenvector can be used to extract the phase lag between the deflection and the pressure field. Both sets are required to represent the response of the coupled system, as demonstrated in Section \ref{sec:dynamics}. 
	
	Since ${\bf \hat B}$ is non-self-adjoint, its right eigenvectors alone do not form a basis with respect to which the system decouples. We therefore introduce the adjoint eigenvalue problem associated with Equation \eqref{eq51}, defined by
	\begin{equation} \label{eq52}
		{\bf \hat B} ^T {\bf \Psi} = \lambda {\bf \Psi}
	\end{equation}
	
	Consequently, the orthogonality of the eigenvectors is expressed through the biorthogonality of the right eigenvectors ${\bf \Phi}_r$ of \eqref{eq51} and the left eigenvectors ${\bf \Psi}_s$ of \eqref{eq52}. Indeed, premultiplying \eqref{eq51} by ${\bf \Psi}_s^T$ and \eqref{eq52}, written for $\lambda_s$, by ${\bf \Phi}_r^T$ and subtracting the two results yields
	\begin{equation} \label{eq53}
		(\lambda_r - \lambda_s)\, {\bf \Psi}_s^T {\bf \Phi}_r = 0
	\end{equation}
	so that ${\bf \Psi}_s^T {\bf \Phi}_r = 0$ whenever $\lambda_r \neq \lambda_s$. The modes can then be scaled to satisfy the biorthonormality conditions \citep{meirovitch2010fundamentals,zezula1977pencil}:
	\begin{align} 
		{\bf \Psi}_s^T {\bf \Phi}_i &= \delta_{rs} \label{eq54} \\
		{\bf \Psi}_s^T {\bf \hat B} {\bf \Phi}_r &= \lambda_r \delta_{rs} \label{eq55}
	\end{align}
	where $\delta$ is the Kronecker delta function. The normalization is applied to one set only, which preserves the conjugate symmetry ${\bf \Psi}_{\bar s} = \bar{\bf \Psi}_s$ of each pair.
	
	The biorthonormality conditions \eqref{eq54} and \eqref{eq55} provide the expansion theorem on which the reduced-order model of Section \ref{sec:rom} is built. To use them, the block structure of the linearized matrices, and hence of ${\bf \hat B}$ and of its eigenvectors, must be made explicit. The linearized stiffness and damping matrices take the form
	\begin{align} 
		&{\bf K}^{(n_w+n_p \times n_w+n_p)}= \left[ \begin{array}{cc}
			{\bf K}_{11}^{(n_w \times n_w)} & {\bf K}_{12}^{(n_w \times n_p)}\\
			{\bf 0}^{(n_p \times n_w)} & {\bf K}_{22}^{(n_p \times n_p)}
		\end{array}\right] \\
		&{\bf C}^{(n_w+n_p \times n_w+n_p)}= \left[ \begin{array}{cc}
			{\bf C}_{11}^{(n_w \times n_w)}&{\bf 0}^{(n_w \times n_p)} \\
			{\bf C}_{21}^{(n_p \times n_w)}&{\bf C}_{22}^{(n_p \times n_p)}
		\end{array}\right]
	\end{align}
	
	The block ${\bf K}_{21}$ vanishes because the linearized pressure contain no term proportional to the displacement (see Equations \eqref{eq37} and \eqref{eq41}), and the block ${\bf C}_{12}$ vanishes because the linearized structural equations contain no time derivative of the pressure (see Equation \eqref{eq36} and \eqref{eq40}). These two zero blocks express the asymmetry of the coupling discussed in Section \ref{sec:formulation}. The blocks ${\bf C}_{11}=\gamma{\bf I}$ and ${\bf C}_{22}$ are diagonal.
	
	The blocks of the matrix $\bf B$ can be expressed as:
	\begin{align*} 
		&{\bf B}_{11}= \left[ \begin{array}{ccc}
			{\bf 0}^{(n_w \times n_w)}&{\bf 0}^{(n_w \times n_p)}&{\bf I}^{(n_w \times n_w)} \\
			{\bf 0}^{(n_p \times n_w)}&{\bf 0}^{(n_p \times n_p)}&{\bf 0}^{(n_p \times n_w)} \\
			-{\bf K}_{11} &-{\bf K}_{12} & -{\bf C}_{11}
		\end{array}\right], \; {\bf B}_{12}= \left[ \begin{array}{c}
			{\bf 0}^{(n_w \times n_p)}\\
			{\bf I}^{(n_p \times n_p)} \\
			{\bf 0}^{(n_w \times n_p)}
		\end{array}\right],\\
		&{\bf B}_{21}= \left[ {\bf 0}^{(n_p \times n_w)} \; -{\bf K}_{22} \; -{\bf C}_{21} \right] \mbox{  and  } {\bf B}_{22}=-{\bf C}_{22}
	\end{align*}
	
	And the matrix $\bf \hat B$ is deduced as follows:
	\begin{equation} \label{eq58}
		{\bf \hat B}= \left[ \begin{array}{ccc}
			{\bf 0}^{(n_w \times n_w)}&{\bf 0}^{(n_w \times n_p)}&{\bf I}^{(n_w \times n_w)} \\
			{\bf 0}^{(n_p \times n_w)}&-{\bf C}_{22}^{-1} {\bf K}_{22}&-{\bf C}_{22}^{-1} {\bf C}_{21} \\
			-{\bf K}_{11} & -{\bf K}_{12} & -{\bf C}_{11}
		\end{array}\right]
	\end{equation}
	where ${\bf C}_{22}$ is invertible, as noted above.
	
	The left eigenvectors are partitioned in the same way as the right ones, is defined as
	\begin{equation*}
		{\bf \Psi} = \left( \begin{array}{c} 
			{\bf \Psi^1}^{(n_w \times 1)} \\
			{\bf \Psi^2}^{(n_p \times 1)} \\ 
			{\bf \Psi^3}^{(n_w \times 1)} 
		\end{array} \right),
	\end{equation*}
	the three blocks being associated with the displacement, pressure and velocity equations respectively.
	
	%%%%%%%%%%%%%%%%%%%%%%%%%%%%%%%%%%%%%%%%%%%%%%%%%%%%%%%%%%%%%%%%%%%%%%%%%%%%%%%%%%%%%%%%%%%%%%%%%%%%%%%%%%%%%%%%%%%%%%%%%%%%%%%%%
	\section{Complex modes of the coupled system}\label{sec:modes}
	This section examines the eigensolutions of the coupled problem. As discussed in the previous section, the complex nature of the eigenvalue problem necessitate a physical description of the eigenvalues and eigenvectors in order to allow their interpretation and then analyze their convergence with respect to the DQM grid.
	
	\subsection{Interpretation of the eigensolutions}
	As discussed above, the eigenvalues of Equation \eqref{eq45} are either complex-conjugate pairs with negative real parts or purely real negative eigenvalues. The first set corresponds to underdamped coupled modes in which the microbeam oscillates and the gas film responds with a phase lag. The second set are modes corresponding to non-oscillatory motions in which a pressure disturbance established in the air film decays, with a relaxation time $\tau_r$, as the gas flows towards the open edges against the viscous resistance of the gap. In this case, the absence of oscillation comes from the absence of inertia in the Reynolds equation.
	
	Several approaches have been proposed to extract real mode shapes \citep{ahmadian1995extracting,balmes1997new} and natural frequencies \citep{lallement1995tutorial} from complex modes. Unlike real modes, complex modes are not energetically independent \citep{ibrahim1999existence}, because of the coupling terms in the damping matrix, motion in one mode dissipates energy in another. Their physical interpretation is accordingly less direct. Since their phase varies from one component to another, complex modes have been interpreted as partially traveling waves, as opposed to the standing waves described by real modes \citep{hoen2005engineering}.
	
	Following references \cite{balmes1997new,hoen2005engineering,lallement1995tutorial}, we define the nondimensional undamped natural frequency and the damping ratio for mode $r$, respectively, as follows:
	\begin{align} 
		& \omega_r = \left| \lambda_r \right| \label{eq59} \\
		&\xi_r= \frac{-\mbox{Re}(\lambda_r)}{\left| \lambda_r \right| } \label{eq60}
	\end{align}
	The corresponding damped natural frequency and modal quality factor can be calculated, respectively, by $\omega_{d,r}=\omega_r \sqrt{1-\xi_r^2}$ and $Q_r=\frac{1}{2 \xi_r}$. 
	
	For the fluid relaxation modes the eigenvalue is real, so that $\xi_r=1$ and $Q_r=1/2$ identically, irrespective of how rapidly the mode decays. These modes are more meaningfully characterized by their relaxation time $\tau_r = 1/|\lambda_r|$, which measures how quickly the film redistributes pressure relative to the structural timescale.
	
	\subsection{Natural frequencies, damping ratios and validation}
	The DQM grid is first fixed by a convergence study, and the resulting modal parameters are then compared with experimental, analytical and numerical
	results from the literature. 
	
	\begin{table}[H]
		\centering
		\caption{Convergence of the coupled eigenvalues with the number of DQM grid points $n$, evaluated at atmospheric pressure for both configurations. $\omega_i$ and $Q_i$ are the nondimensional undamped frequency and the quality factor defined by Equations \eqref{eq59} and \eqref{eq60}. S, A, C, and F denote symmetric, antisymmetric, coupled modes, and fluid relaxation mode, respectively. The relaxation time $\tau_i = 1/|\lambda_i|$ is reported for fluid modes.}
		\label{tab5}
		\begin{tabular}{llccccc}
			\hline
			\multicolumn{7}{c}{CC microbeam, $p_a=1.01325\times10^{5}\,$Pa, and $V_{DC}=2$\,V}     \\
			$n$               &            &    $9$    &   $11$    &   $13$    &   $15$    &   $17$    \\
			$(n_w,n_p)$       &            &  (7,28)   &  (9,45)   &  (11,66)  &  (13,91)  & (15,120)  \\ \hline
			1$^{st}$ C mode S & $\omega_i$ & 123.34507 & 123.34455 & 123.34461 & 123.34464 & 123.34464 \\
			& $Q_i$      &  4.24462  &  4.24448  &  4.24452  &  4.24451  &  4.24451  \\
			1$^{st}$ C mode A & $\omega_i$ & 137.83497 & 137.83275 & 137.83262 & 137.83285 & 137.83289 \\
			& $Q_i$      &  4.72413  &  4.72416  &  4.72452  &  4.72456  &  4.72455  \\
			2$^{nd}$ C mode S & $\omega_i$ & 174.45063 & 174.92862 & 174.93284 & 174.93353 & 174.93366 \\
			& $Q_i$      &  7.18533  &  7.23011  &  7.23267  &  7.23290  &  7.23286  \\
			1$^{st}$ F mode S & $\omega_i$ &  1.09793  &  1.09790  &  1.09790  &  1.09790  &  1.09790  \\
			& $\tau_i$   &  0.91080  &  0.91083  &  0.91083  &  0.91083  &  0.91083  \\ \hline
			\multicolumn{7}{c}{CF microbeam, $p_a=1.01325\times10^{5}\,$Pa, and $V_{DC}=0$\,V}     \\
			$n$               &            &    $9$    &   $13$    &   $17$    &   $21$    &   $25$    \\
			$(n_w,n_p)$       &            &  (7,28)   &  (11,66)  & (15,120)  & (19,190)  & (23,276)  \\ \hline
			1$^{st}$ C mode   & $\omega_i$ &  3.53401  &  3.53372  &  3.53375  &  3.53375  &  3.53376  \\
			& $Q_i$      &  1.07859  &  1.10276  &  1.10619  &  1.10711  &  1.10757  \\
			2$^{nd}$ C mode   & $\omega_i$ & 22.14531  & 22.14230  & 22.14226  & 22.14220  & 22.14223  \\
			& $Q_i$      &  6.79196  &  6.94432  &  6.96549  &  6.97193  &  6.97484  \\
			3$^{rd}$ C mode   & $\omega_i$ & 62.07705  & 61.98668  & 61.98680  & 61.98653  & 61.98665  \\
			& $Q_i$      & 19.82055  & 20.23473  & 20.30133  & 20.31953  & 20.32845  \\
			1$^{st}$ F mode   & $\omega_i$ & 270.26302 & 303.24912 & 303.24918 & 303.24918 & 303.24918 \\
			& $\tau_i$   & 0.003700  & 0.003298  & 0.003298  & 0.003298  & 0.003298  \\ \hline
		\end{tabular}
	\end{table}
	
	Table \ref{tab5} reports the convergence of the coupled eigenvalues with the number of DQM grid points using Equations \eqref{eq59} and \eqref{eq60}. Both CF and CC configurations are evaluated at atmospheric pressure, supposed to be the most sensitive case. First, the quality factors converge more slowly than the frequencies in every case, which is expected because $Q_i$ depends on the real part of the eigenvalue and therefore on the resolution of the fluid partition, whereas the frequency is governed mainly by the structure. Second, the CF microbeam requires a finer grid than the CC one. Its quality factors still vary by about $0.05$\% between $n=21$ and $n=25$, whereas those of the CC microbeam vary by less than $0.003$\% from $n=13$ onwards. The fluid relaxation mode is the slowest to converge in frequency, for the CF microbeam at $n=9$ it is in error by $11$\%, while the coupled modes are already accurate to within $0.2$\%. This difference originates in the boundary conditions of the film. The CC film is sealed at both ends so the gas escapes only laterally through the vented edges, whereas the CF film is additionally opened at the free end. It is therefore the fluid that governs the grid requirement in the coupled problem. In what follows, $n=21$ is used for both configurations.
	
	To validate the methodology, we compare the natural frequencies, damping ratios and quality factors results reported in Table \ref{tab2} for the CC microbeam and in Table \ref{tab3} for the CF microbeam with those given by references \citep{tilmans1994electrostatically,nayfeh2004squeeze,pandey2007effect,chaterjee2012squeeze}. The CC microbeam is studied at $p_a=6\times10^{-4}$\,Pa and $V_{DC}=2$\,V. The experimental data available for this microbeam concern the symmetric modes only, since the antisymmetric modes are not excited by the uniform electrostatic actuation used in the reference works. For the CC configuration, the nondimensional structural damping coefficient $\gamma=0.0014515$ is deduced from the measured total quality factor of the fundamental mode reported by \citet{tilmans1994electrostatically}. The CF microbeam is studied without DC voltage at atmospheric pressure. The calculated natural frequencies and damping ratios due to squeeze film damping are in very good agreement with those reported in the references.
	
	\begin{table}[H]
		\centering
		\caption{Validation of the calculated symmetric mode natural frequencies for the CC microbeam $p_a=6\times 10^{-4}$Pa and $V_{DC}=2$V.}
		\label{tab2}
		\begin{tabular}{lcccc}
			&  $\omega_1/2\pi$  &  $\omega_2/2\pi$  & $Q_1$ & $Q_2$ \\
			&       (kHz)       &       (kHz)       &       &       \\ \hline
			Experimental \citep{tilmans1994electrostatically} & 163.215$\pm3.962$ & 762.350$\pm$4.729 & 18500 &   -   \\
			FEM \citep{nayfeh2004squeeze}                     &      161.011      &         -         & 18500 &   -   \\
			Present model                                     &      162.553      &      775.876      & 18500 & 88329 \\ \hline
		\end{tabular}
	\end{table}
	
	\begin{table}[H]
		\centering
		\caption{Validation of the calculated natural frequencies for the CF microbeam $p_a=1.01325\times10^{5}\,$Pa and $V_{DC}=0$V.}
		\label{tab3}
		\begin{tabular}{lcccccc}
			& $\omega_1/2\pi$ & $\omega_2/2\pi$ & $\omega_3/2\pi$ & $\xi_1$ & $\xi_2$ & $\xi_3$ \\
			&      (kHz)      &      (kHz)      &      (kHz)      &         &         &         \\ \hline
			Experimental \citep{pandey2007effect}   &       43        &       245       &       690       &  0.415  &  0.066  &  0.027  \\
			Analytical \citep{pandey2007effect}     &      44.8       &      280.8      &      786.3      &  0.422  &  0.067  &  0.024  \\
			FEM \citep{pandey2007effect}            &        -        &        -        &        -        &  0.450  &  0.072  &  0.025  \\
			Analytical \citep{chaterjee2012squeeze} &      43.81      &        -        &        -        & 0.4484  & 0.0713  & 0.0247  \\
			FEM \citep{chaterjee2012squeeze}        &        -        &        -        &        -        & 0.4475  & 0.0710  & 0.0244  \\
			Present model                           &      43.93      &     275.27      &     770.61      & 0.4515  & 0.0717  & 0.0246  \\ \hline
		\end{tabular}
	\end{table}
	
\begin{figure}[H] 
		\centering
		\includegraphics[width=0.7\textwidth]{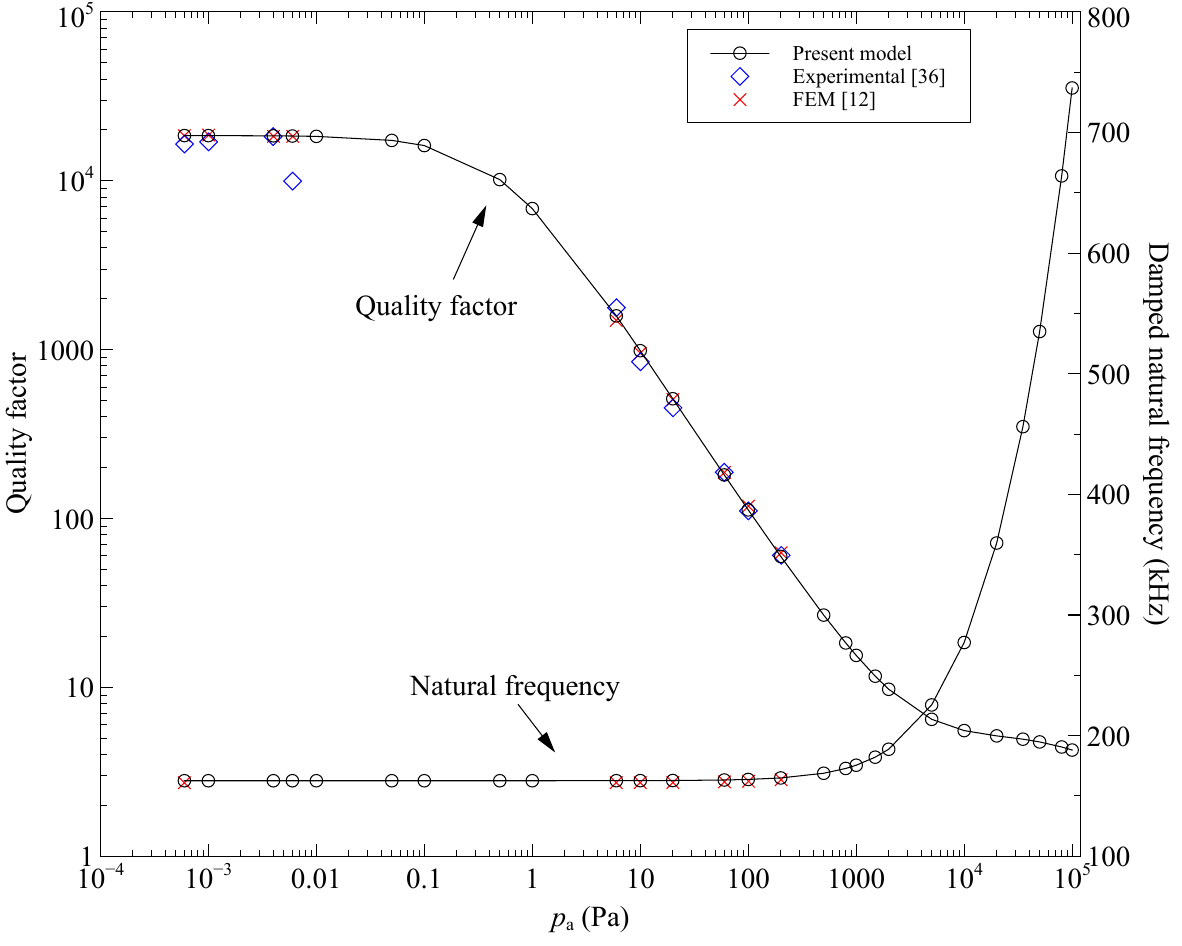}
		\caption{Effect of the ambient pressure on the quality factor (left axis) and the damped natural frequency (right axis) of the first symmetric mode of the CC microbeam, compared with the experimental data of Tilmans and Legtenberg \citep{tilmans1994electrostatically} and the finite element results of Nayfeh and Younis \citep{nayfeh2004squeeze}. $V_{DC}=2\,$V} 
		\label{fig3}
	\end{figure}

\begin{figure}[H] 
		\centering
		\includegraphics[width=0.7\textwidth]{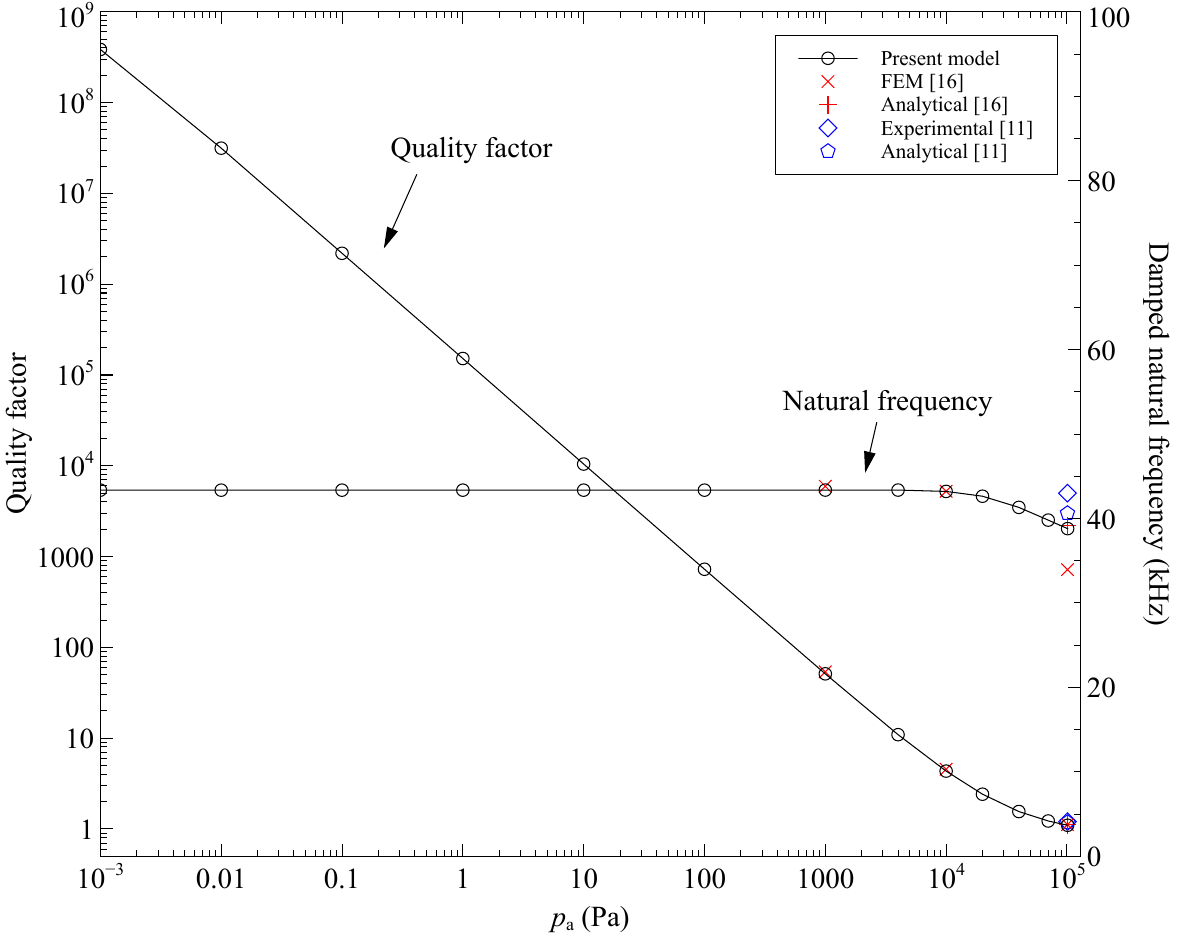}
		\caption{Effect of the ambient pressure on the quality factor (left axis) and the damped natural frequency (right axis) of the first mode of the CF microbeam, compared with the experimental, analytical and finite element results of Pandey and Pratap \citep{pandey2007effect} and Chaterjee and Pohit \citep{chaterjee2012squeeze}.} 
		\label{fig4}
	\end{figure}

	Figures \ref{fig3} and \ref{fig4} show the influence of the ambient pressure on the first damped natural frequency and the associated quality factor, over the whole range of pressure regimes, and the calculated values agree with the available experimental and numerical results. For the CC microbeam the quality factor falls by more than three orders of magnitude between the molecular and continuum regimes, while the frequency of the first coupled mode remains constant up to about $10^2$\,Pa and then increases sharply towards atmospheric pressure. This increase reflects the elastic contribution of the air film that cannot escape within a period and acts as a spring. In this range the first coupled mode is no longer the mode with the lowest eigenvalue, fluid relaxation modes with zero damped frequency appearing below it. For the CF microbeam this stiffening is absent. The film, open at the free end as well as along the lateral edges, relaxes rapidly, and the frequency decreases slightly at high pressure, indicating the predominance of the viscous effect.
	
	At the highest pressures, the stiffening is so strong that two of the assumptions that the Reynolds equation is based on are no longer true. At $p_a=1.01325\times10^{5}$\,Pa the predicted frequency corresponds to a squeeze Reynolds number $\rho_g \hat{\omega} d^2/\mu \approx 0.4$, where $\rho_g$ is the air density and $\hat\omega$ the frequency of the motion, so that gas inertia is no longer strictly negligible, and to an acoustic wavelength of the same order as the microbeam length, so that wave propagation within the film, which the lubrication model cannot represent, is no longer excluded. The predicted frequencies at the highest pressures should therefore be regarded as indicative of the trend rather than quantitatively exact. The comparisons with experimental data in Table \ref{tab2} and Figure \ref{fig3} lie below $2\times10^{2}$\,Pa, where both conditions are satisfied.
	
	Table \ref{tab4} lists the frequencies, damping ratios and quality factors of the lowest modes of both microbeams, identified by their nature and, for the CC microbeam, by their parity. The coupled modes occur as complex-conjugate pairs sharing the same frequency, and the fluid modes as real eigenvalues. The composition of the two spectra differs. For the CC microbeam the fluid modes are mixed with the coupled ones, and the antisymmetric coupled modes alternate with the symmetric ones. For the CF microbeam the six lowest modes are all coupled and the fluid modes are all above $\omega \approx 303$. This difference persists at equal pressure, the first fluid mode lying at $\omega=1.10$ for the CC microbeam against $303.25$ for the CF microbeam at atmospheric pressure (Table \ref{tab5}), and it reflects the venting of the CF film at its free end, which allows the pressure to relax far more rapidly than the CC film.
	
	\begin{table}[H]
	\centering
	\caption{Nondimensional undamped frequencies, damping ratios and quality factors for different mode shapes ($n=21$), for the CC and CF microbeams. S, A, C and F denote symmetric, antisymmetric, coupled, and fluid relaxation modes.}
	\label{tab4}
	\begin{tabular}{cccccc}
		\multicolumn{6}{c}{CC microbeam, $p_a=6 \times 10^{-4}\,$Pa, and $V_{DC}=2\,$V}  \\
		Mode  & $\omega_i$ &        $\xi_i$        & $Q_i$  & Nature &      Parity       \\
		 1-2  &   26.866   & 2.70$\times$10$^{-5}$ & 18500  &   C    &         S         \\
		 3-4  &   68.184   & 1.06$\times$10$^{-5}$ & 46960  &   C    &         A         \\
		  5   &  104.354   &          --           &   --   &   F    &         S         \\
		  6   &  115.370   &          --           &   --   &   F    &         A         \\
		 7-8  &  128.235   & 5.66$\times$10$^{-6}$ & 88329  &   C    &         S         \\
		  9   &  147.794   &          --           &   --   &   F    &         S         \\
		 10   &  202.082   &          --           &   --   &   F    &         A         \\
		11-12 &  207.647   & 3.49$\times$10$^{-6}$ & 143039 &   C    &         A         \\ \hline
		\multicolumn{6}{c}{CF microbeam, $p_a=1.01325\times10^5\,$Pa, and $V_{DC}=0\,$V} \\
		Mode  & $\omega_i$ &        $\xi_i$        & $Q_i$  & Nature &      Parity       \\
		 1-2  &   3.534    &        0.4515         &  1.11  &   C    &        --         \\
		 3-4  &   22.142   &        0.0717         &  6.97  &   C    &        --         \\
		 5-6  &   61.987   &        0.0246         & 20.33  &   C    &        --         \\
		 7-8  &  121.401   &        0.0113         & 44.41  &   C    &        --         \\
		9-10  &  200.515   &        0.0056         & 89.50  &   C    &        --         \\
		11-12 &  299.267   &        0.0028         & 176.18 &   C    &        --         \\
		 13   &  303.249   &          --           &   --   &   F    &        --         \\
		 14   &  305.863   &          --           &   --   &   F    &        --         \\
		 15   &  310.845   &          --           &   --   &   F    &        --         \\
		 16   &  318.337   &          --           &   --   &   F    &        --         \\ \hline
	\end{tabular}
\end{table}	
	
	\subsection{Equivalent real mode shapes}
	A complex eigenvector represents a spatial shape whose phase may vary from one degree of freedom to another. Although the complex modes are used directly as the projection basis in Section \ref{sec:rom}, it is useful to associate with each of them an equivalent real mode shape, both for comparison with classical undamped modal analysis and for visualization. 
	
	The construction take into account the fact that an eigenvector is defined only up to an arbitrary complex factor. So that it is possible to determine the rotation that renders the eigenvector as close to real as possible in the least-squares sense.

	Let ${\bf \Phi}_r$ be a complex right eigenvector and ${\bf \Phi}^{\bf{p}}_r$ one of its partitions, ${\bf \Phi}^{\bf{1}}_r$ denoting the displacement and ${\bf \Phi}^{\bf{2}}_r$ the pressure. The rotation is applied to each partition separately because they carry different physical quantities. The present construction therefore measures the complexity internal to each field. 
	
	Consider the rotated partition $e^{\mathrm{i}\alpha}{\bf \Phi}^{\bf{p}}_r = {\bf a}_r(\alpha) + \mathrm{i}\,{\bf b}_r(\alpha)$, with ${\bf a}_r$ and ${\bf b}_r$ real. Its real part is
	\begin{equation}
		\|{\bf a}_r(\alpha)\|^2 = \tfrac12\left( {{\bf \Phi}^{\bf{p}}_r}^H{\bf \Phi}^{\bf{p}}_r + \left|{{\bf \Phi}^{\bf{p}}_r}^T{\bf \Phi}^{\bf{p}}_r\right| 	\cos\left(2\alpha + \arg\left({{\bf \Phi}^{\bf{p}}_r}^T{\bf \Phi}^{\bf{p}}_r\right)\right)\right)
	\end{equation}
	which is maximal for
	\begin{equation} \label{eq61}
		\alpha^{{\bf{p}}}_r = -\frac{1}{2} \arg \left( {{\bf \Phi}^{\bf{p}}_r}^T {\bf \Phi}^{\bf{p}}_r \right)
	\end{equation}
	
	The stationary points of $\|{\bf a}_r\|^2$ are those for which ${\bf a}_r \cdot {\bf b}_r = 0$. They occur every $\pi/2$, alternately maxima and minima, so that Equation \eqref{eq61} defines the maximum up to a multiple of $\pi$, which only changes the sign of the real shape. The equivalent real mode of the partition is then
	\begin{equation} \label{eq62}
		{\bf \Phi}^{{\bf{p}},R}_r = \mathrm{Re}\left( e^{\mathrm{i}\alpha^{{\bf{p}}}_r} {\bf \Phi}^{\bf{p}}_r \right) = {\bf a}_r
	\end{equation}
	
	The residual imaginary part provides a natural measure of how far the partition is different from a real shape,
	\begin{equation} \label{eq63}
		\epsilon^{{\bf{p}}}_r = \frac{\| {\bf b}_r \|}{\| {\bf a}_r \|}
	\end{equation}
	which vanishes when the partition is real up to a global phase, that is when all its components reach their extrema at the same instant.
	
	The consistency between the equivalent real shapes and the complex partitions from which they are derived is assessed using the modal assurance criterion (MAC) \citep{allemang2003modal}, evaluated as a cross-correlation between the two sets,
	\begin{equation} \label{eq64}
		\mathrm{MAC}^{{\bf{p}}}_{rs} = \frac{\left| {{\bf \Phi}^{{\bf{p}},R}_r}^{T} \overline{{\bf \Phi}^{{\bf{p}}}_s} \right|^2} 	{\left( {{\bf \Phi}^{{\bf{p}},R}_r}^{T} {\bf \Phi}^{{\bf{p}},R}_r \right) \left( {{\bf \Phi}^{{\bf{p}}}_s}^{H} {\bf \Phi}^{{\bf{p}}}_s \right)}
	\end{equation}
	where the superscript $H$ denotes the conjugate transpose. Since the two sets being correlated are distinct, the resulting matrix is not symmetric. The matrices obtained for the displacement and pressure partitions are denoted $\mathrm{MAC}^{c}$ and $\mathrm{MAC}^{f}$.
	
	The diagonal entries of Equation \eqref{eq64} admit a simple interpretation. Since the rotation \eqref{eq61} makes ${\bf a}_r$ and ${\bf b}_r$ orthogonal, the diagonal reduces to
	\begin{equation} \label{eq65}
		\mathrm{MAC}^{{\bf{p}}}_{rr} = \frac{1}{1+\left(\epsilon^{{\bf{p}}}_r\right)^2} = \frac{{{\bf \Phi}^{{\bf{p}}}_r}^H{\bf \Phi}^{{\bf{p}}}_r + \left|{{\bf \Phi}^{{\bf{p}}}_r}^T{\bf \Phi}^{{\bf{p}}}_r\right|} {2\,{{\bf \Phi}^{{\bf{p}}}_r}^H{\bf \Phi}^{{\bf{p}}}_r}
	\end{equation}
	so that the diagonal of each MAC matrix is itself a measure of the complexity of the corresponding field, which can moreover be computed directly from the eigenvector without constructing the rotation. It lies between $\frac12$ and $1$. A value of unity identifies a field that is real up to a global phase, whereas the lower bound is reached when ${{\bf \Phi}^{{\bf{p}}}_r}^T{\bf \Phi}^{{\bf{p}}}_r = 0$, for a field whose components are distributed uniformly in phase, similar to a traveling wave.
	
	The modes are classified according to the nature of their eigenvalue, by defining the sets of coupled and fluid modes
	$$
	\mathcal{I}_c=\{\,r:\ \operatorname{Im}\lambda_r>0\,\},
	\qquad
	\mathcal{I}_f=\{\,r:\ \operatorname{Im}\lambda_r=0\,\}.
	$$
	The structural matrix is restricted to $\mathcal{I}_c$, since the fluid relaxation modes have essentially no structural content. The fluid matrix spans $\mathcal{I}_c\cup\mathcal{I}_f$, sorted by increasing $|\lambda_r|$, since the pressure field is set in motion both by the coupled modes, through the squeeze motion of the structure, and by the fluid relaxation modes. A real eigenvalue of a real matrix having a real eigenvector, therefore the diagonal elements of the fluid relaxation modes are equal to unity. The information carried by the fluid matrix therefore lies in the diagonal entries of the coupled modes and in the off-diagonal entries.
	
\begin{figure}[H]	
		\centering
		\subfloat[][CC microbeam, $p_a=6\times10^{-4}$\,Pa.\label{fig5a}]{\scalebox{1}{\includegraphics[width=0.7\textwidth]{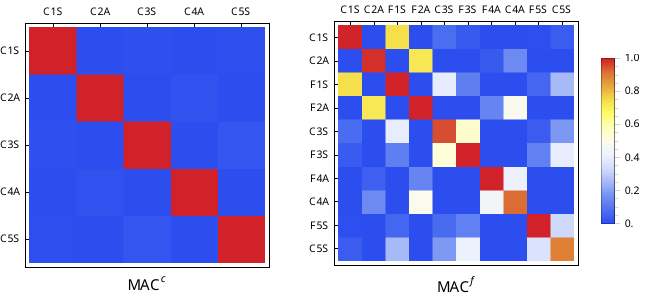}}}\\
		\subfloat[][CF microbeam, $p_a=1.01\times10^{5}$\,Pa.\label{fig5b}]{\scalebox{1}{\includegraphics[width=0.7\textwidth]{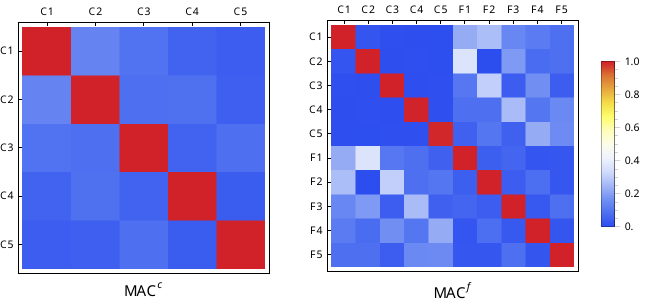}}}\\
		\subfloat[][CC microbeam, $p_a=1.01\times10^{5}$\,Pa.\label{fig5c}]	{\scalebox{1}{\includegraphics[width=0.7\textwidth]{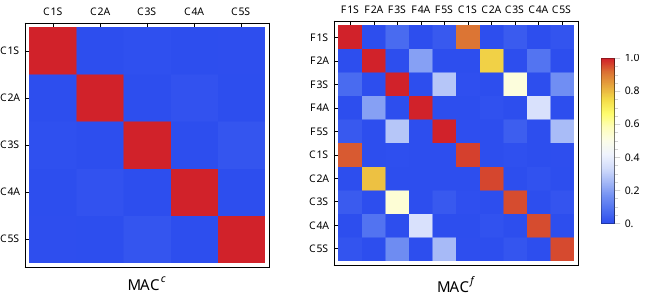}}}
		\caption{MAC matrix between the equivalent real mode shapes and the corresponding complex eigenvectors, for the CC and CF microbeams.}
		\label{fig5}
	\end{figure}

	The two matrices are reported in Figure \ref{fig5} for the lowest modes of each set. The structural partitions are real to a very good approximation in every case. All coupled modes satisfy $\mathrm{MAC}^{s}_{rr} > 0.99$, for both microbeams and at both pressures. The microbeam therefore moves synchronously in each coupled mode, every point reaching its extremum at the same instant, even when the squeeze-film effect is strong.
	
	The pressure partitions behave differently. For the CC microbeam at $p_a = 6\times10^{-4}$\,Pa, only one coupled mode satisfies $\mathrm{MAC}^{f}_{rr} > 0.99$, the diagonal decreasing with the mode order from $0.998$ for the fundamental mode to $0.896$ for the fifth one (Figure \ref{fig5a}). At atmospheric pressure no coupled mode satisfies the criterion, the five lowest ones lying between $0.956$ and $0.969$ (Figure \ref{fig5c}). When the number of grid points is increased from $n=21$ to $31$ and $43$, these values change by less than $0.3$\%, while the structural values are unchanged to fifth digit.
	
	For the CF microbeam at atmospheric pressure six coupled modes satisfy the criterion, the five lowest ones exceeding $0.995$ (Figure \ref{fig5b}). Thanks to the additional opening at the free end, the pressure follows the deflection almost quasi-statically and its distribution remains real. For the CC microbeam the film is sealed at the ends and the coupled modes therefore cannot be represented by real shapes, not because of their structural deformation which are synchronous, but because of their pressure field, whose phase varies over the film. 
	
	The off-diagonal entries of the fluid matrix reveal how these pressure fields are composed. For the CC microbeam, each coupled mode correlates with the fluid relaxation mode of the same parity and of the same order. At atmospheric pressure the correlation reaches $0.94$ between the first coupled and the first fluid mode and $0.79$ for the second pair, decreasing to $0.26$ for the fifth. The pressure field of the lowest coupled modes is largely similar to the corresponding film relaxation mode. At $p_a = 6\times10^{-4}$\,Pa the pairing persists but decays more slowly with the order, the correlation decreasing from $0.74$ for the first pair to $0.36$ for the fifth. For the CF microbeam the correlations are weaker, and shifted by one order. The pressure field of the $r^{th}$ coupled mode correlates with the $(r-1)^{th}$ fluid relaxation mode, the correlation decreasing from $0.36$ for the second coupled mode to $0.22$ for the fifth. The phase between the pressure and the deflection is examined in Section \ref{sec:phase}.
	
\begin{figure}[H]	
		\centering
		\subfloat[][CC microbeam.]{\scalebox{0.95}{\includegraphics[width=0.5\textwidth]{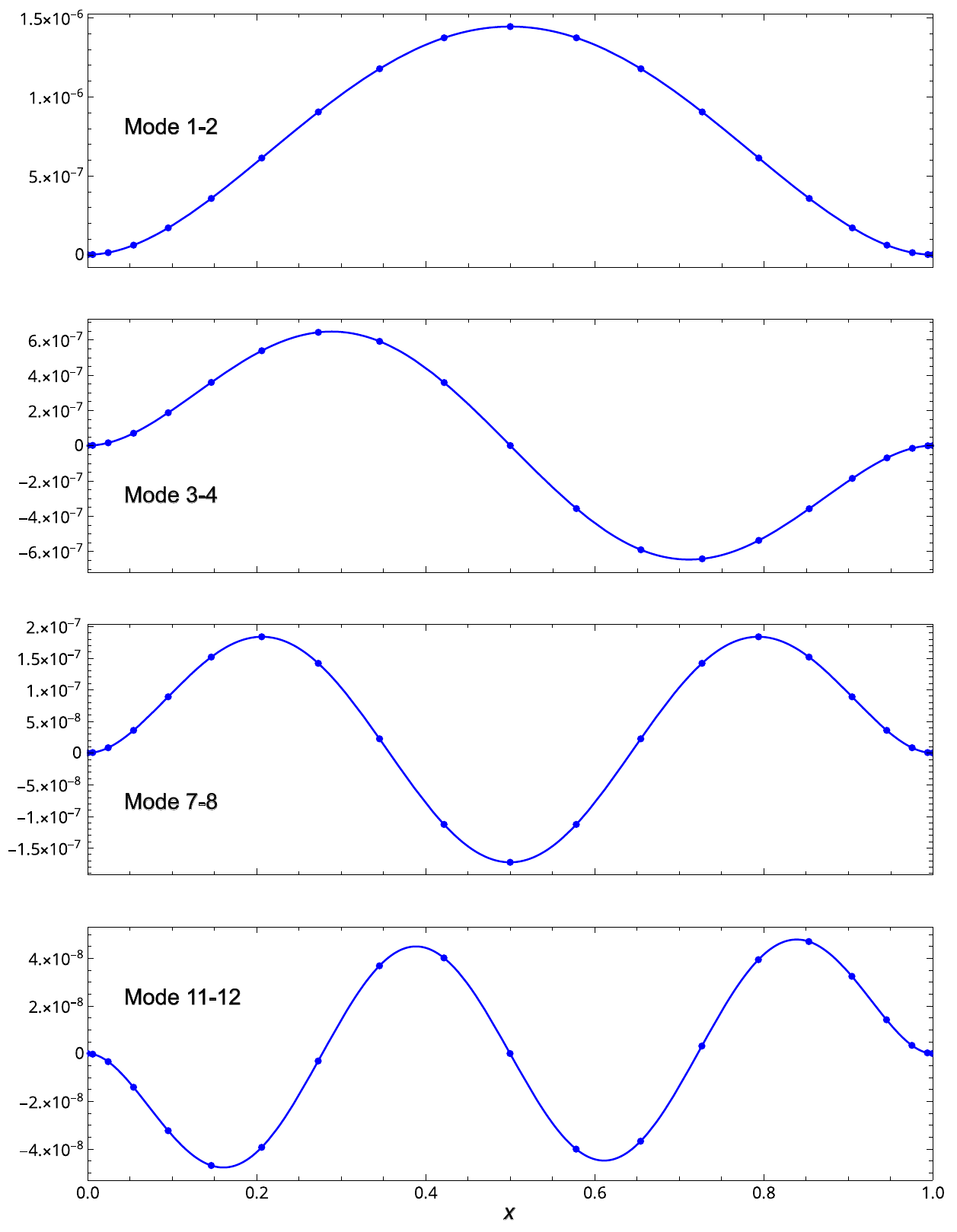}}} \quad
		\subfloat[][CF microbeam.]{\scalebox{0.95}{\includegraphics[width=0.5\textwidth]{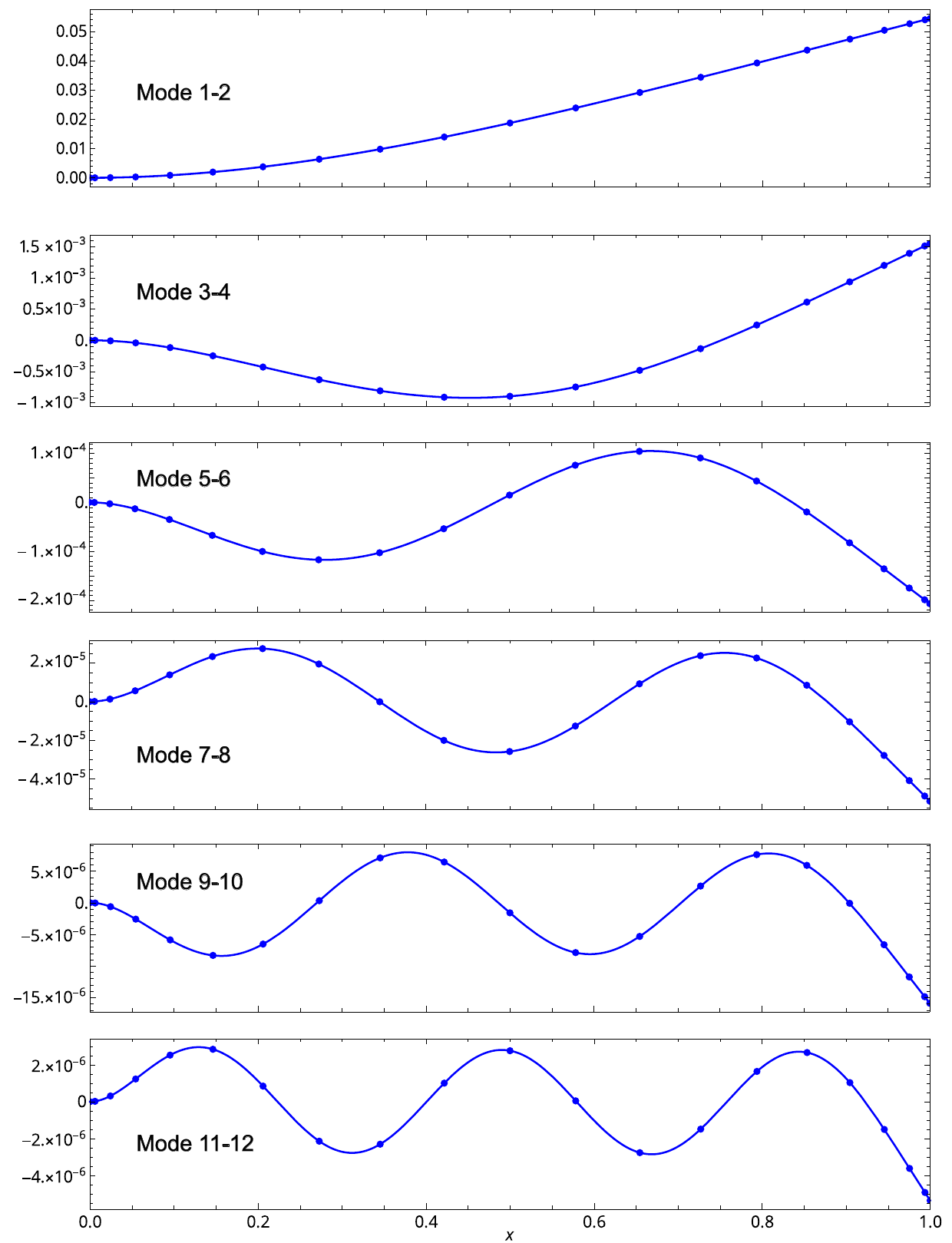}}}
		\caption{Equivalent real structural mode shapes along the microbeam reconstructed from the complex eigenvectors of the coupled problem.}
		\label{fig6}
	\end{figure}

	The equivalent real structural mode shapes are shown in Figure \ref{fig6}. For the CC microbeam, the four real shapes plotted correspond to the eight complex eigenvectors of the coupled modes of Table \ref{tab4}, and for the CF microbeam the six real shapes correspond to its first twelve eigenvectors. In both cases the shapes are consistent with those of the corresponding undamped structural problem.
	
\begin{figure}[H] 
		\centering
		\includegraphics[width=0.9\textwidth]{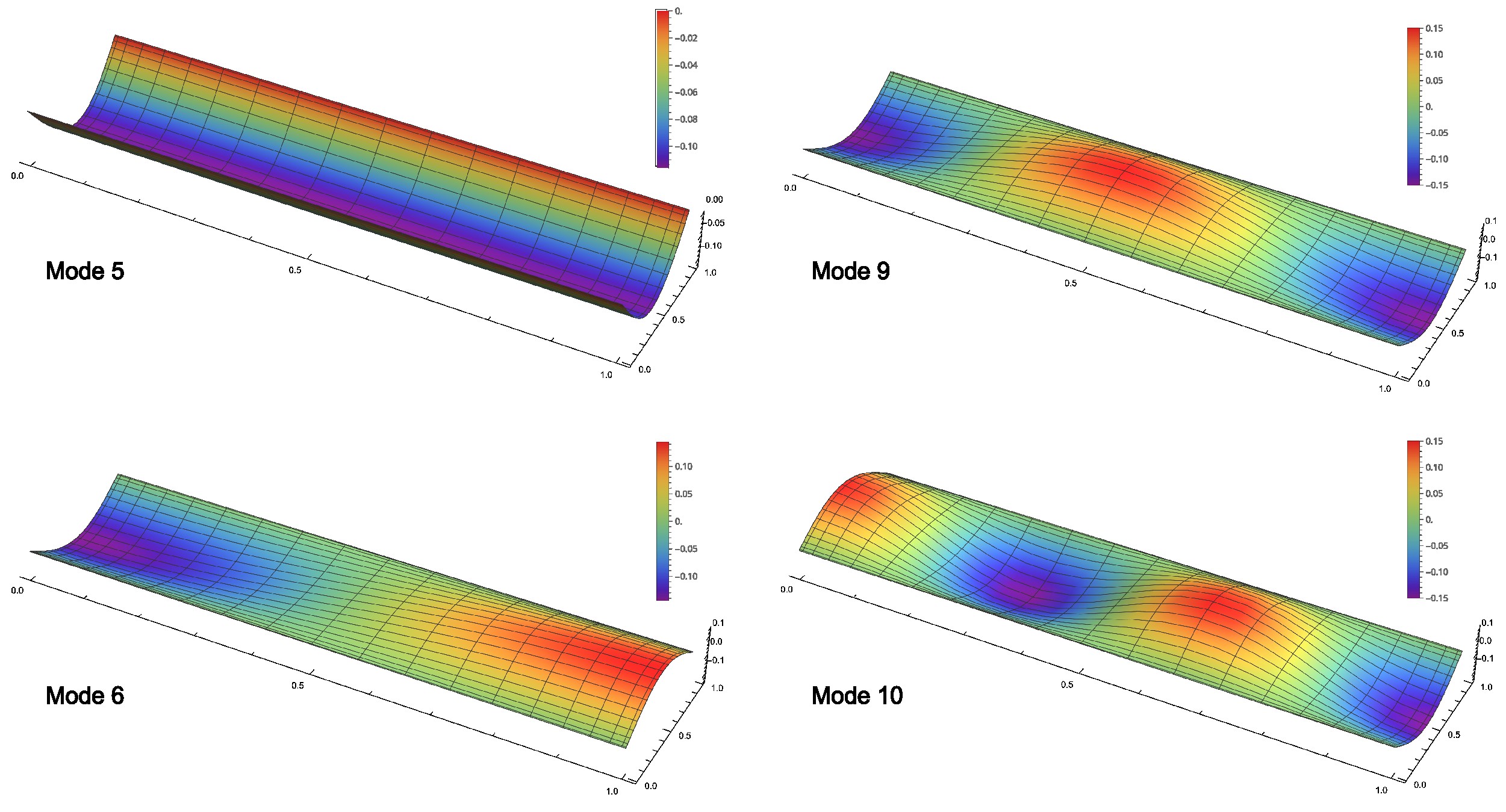}
		\caption{Equivalent real fluid mode shapes over the air film of the CC microbeam. Mode numbers refer to Table \ref{tab4}.} 
		\label{fig7}
	\end{figure}

\begin{figure}[H] 
		\centering
		\includegraphics[width=0.9\textwidth]{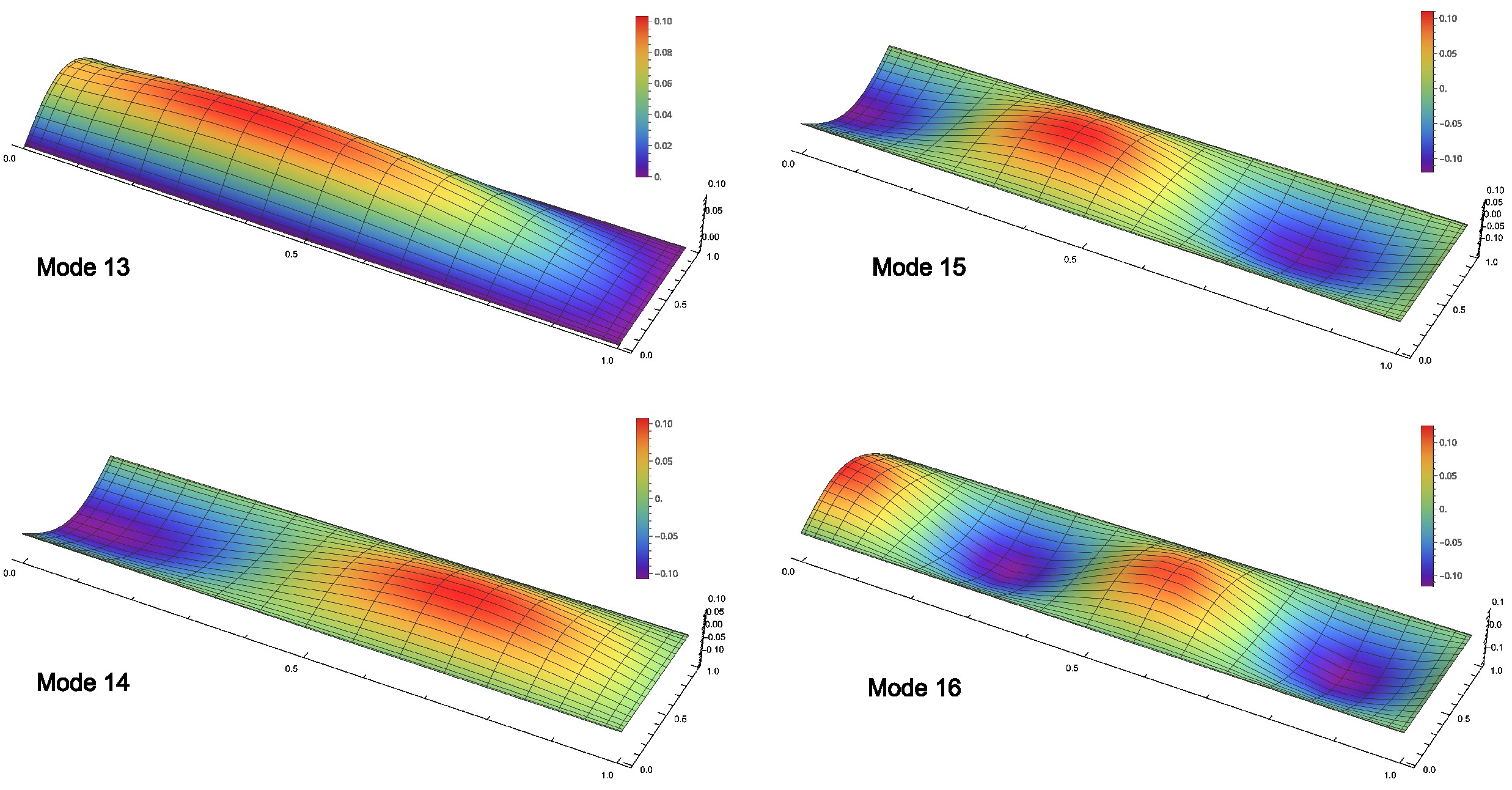}
		\caption{Equivalent real fluid mode shapes over the air film of the CF microbeam. Mode numbers refer to Table \ref{tab4}.} 
		\label{fig8}
	\end{figure}

	The pressure partitions of the modes of Table \ref{tab4} are shown in Figure \ref{fig7} for the CC microbeam and in Figure \ref{fig8} for the CF microbeam. They fall into two categories. The pressure fields of the coupled modes, identified as C in Table \ref{tab4}, follow the deflection of the corresponding structural mode, the film being squeezed where the microbeam moves. The pressure fields of the fluid relaxation modes, identified as F, describe redistributions of the gas with negligible structural participation. Their shapes alternate between regions of high and low relative pressure.
	
	
	\subsection{Phase content of the coupled complex modes} \label{sec:phase}
	The information carried by a complex mode that a real mode cannot represent is the relative phase between its components. Writing the $r^{th}$ right eigenvector in polar form, ${\bf \Phi}_r = |{\bf \Phi}_r|\, e^{\mathrm{i}\theta_r}$, the modulus gives the amplitude of each degree of freedom while $\theta_r$ fixes the instant at which it reaches its extremum. Since an eigenvector is defined up to an arbitrary complex factor, each coupled mode is characterized by the phase of the pressure field partition relative to the deflection,
	\begin{equation} \label{eq66}
		\Delta\theta_r(x,y) = \arg\left( {\bf \Phi}^{\bf 2}_r(x,y)\,
		\overline{{\bf \Phi}^{\bf 1}_r(x)} \right)
	\end{equation}
	
	In Equation \eqref{eq66} a positive $\Delta\theta_r$ indicates that the pressure leads the deflection. When $\Delta\theta_r = 0$ the air film acts as a pure spring and when $\Delta\theta_r=\pi/2$ the pressure is in phase with the velocity and the film acts as a pure damper. At each point of the film, the pressure of a coupled mode oscillates at the frequency of the mode, and it can be separated into two parts: one varies proportionally to the deflection with a relative amplitude $\cos\Delta\theta_r$ and therefore acts as a spring, and the other proportionally to the velocity with a relative amplitude $\sin\Delta\theta_r$ and acts as a damper. When $\cos\Delta\theta_r$ is positive, the pressure rises as the gap closes so that the film stiffens the microbeam. When it is negative, the pressure falls as the gap closes and draws the microbeam further towards the substrate, so that the film softens it. For the second part, when $\sin\Delta\theta_r$ is positive, the force opposes the motion and the film dissipates energy, whereas a negative value corresponds to a region of the film returning energy to the structure. However, it should be noted that this decomposition is local and assumes a harmonic motion. The effective stiffness and damping contributed by the film to a mode result from its integration over the film, weighted by the mode shape, so that regions of different type coexist within the same mode. It is this mixed character that no real mode and no single lumped damping coefficient can represent.
	
	The phase of the structural partition along the microbeam, $\arg {\bf \Phi}^{\bf 1}_r(x)$, indicates whether the microbeam moves synchronously. For the fundamental coupled mode of the CC microbeam at $p_a = 6\times10^{-4}$\,Pa it is uniform along the microbeam, similar to a classically damped system. At atmospheric pressure it varies by about $0.12$\,rad, following a distribution that peaks at midspan. For the CF microbeam at atmospheric pressure it is uniform to within $8\times10^{-3}$\,rad.
	
\begin{figure}[H]	
		\centering
		\subfloat[][$\Delta\theta_1$ for CC at $p_a=6\times10^{-4}$\,Pa.\label{fig9a}]{\scalebox{0.95}{\includegraphics[width=0.5\textwidth]{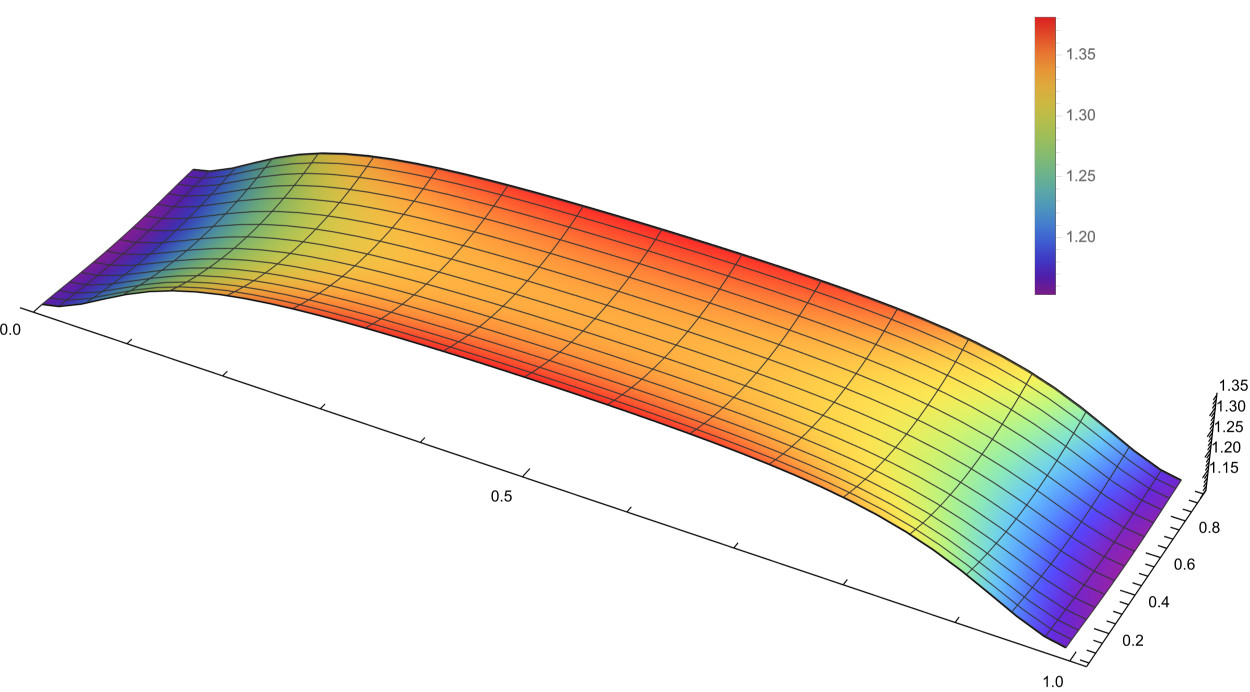}}} \quad
		\subfloat[][$D_1$ for CC at $p_a=6\times10^{-4}$\,Pa.\label{fig9b}]{\scalebox{0.95}{\includegraphics[width=0.5\textwidth]{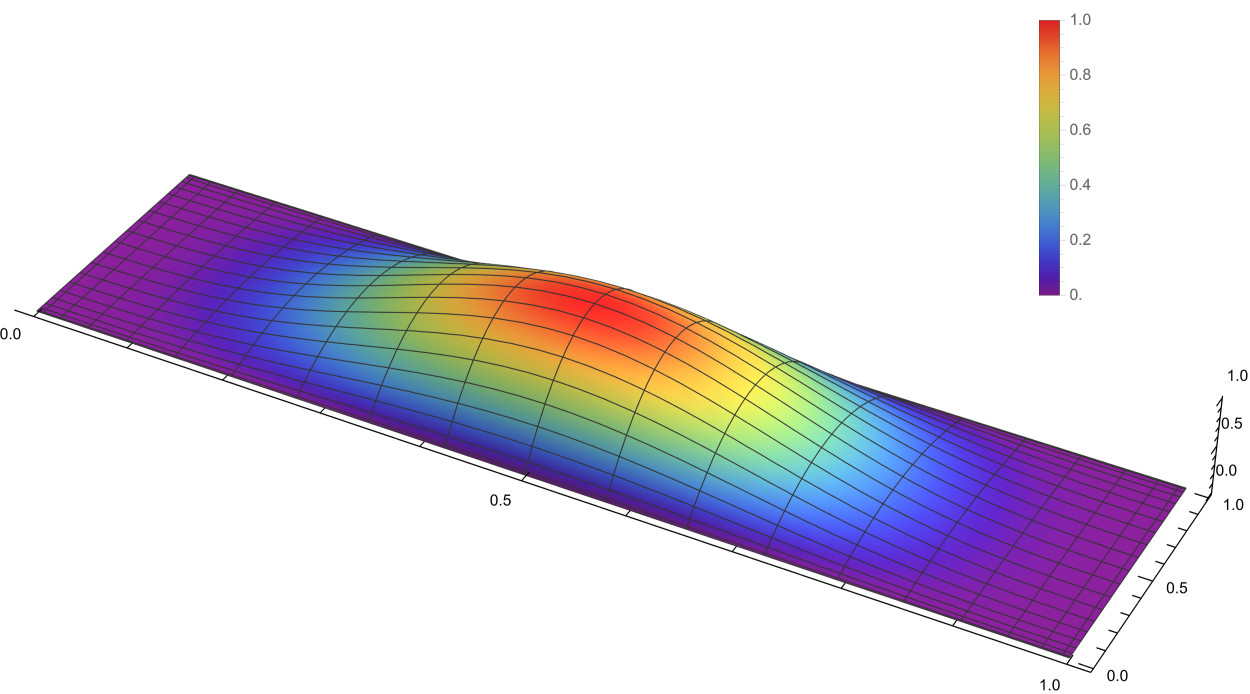}}}\\
		\subfloat[][$\Delta\theta_1$ for CC at $p_a=1.01\times10^{5}$\,Pa.\label{fig9c}]{\scalebox{0.95}{\includegraphics[width=0.5\textwidth]{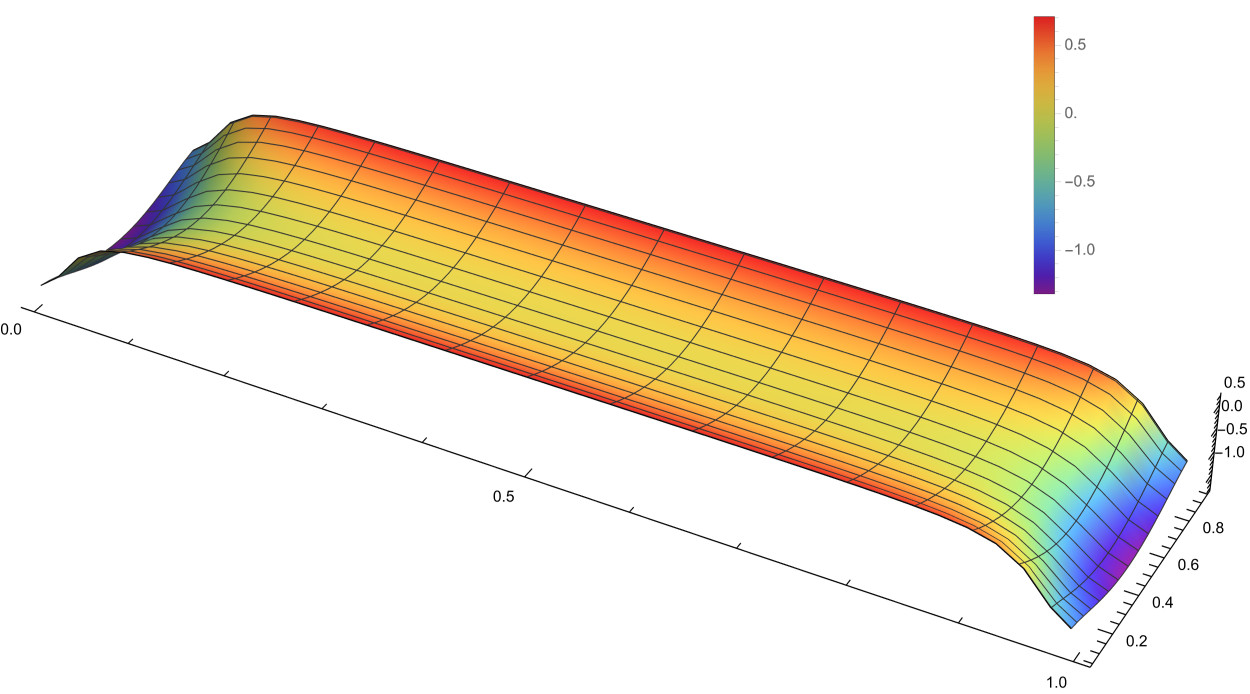}}} \quad
		\subfloat[][$D_1$ for CC at $p_a=1.01\times10^{5}$\,Pa.\label{fig9d}]{\scalebox{0.95}{\includegraphics[width=0.5\textwidth]{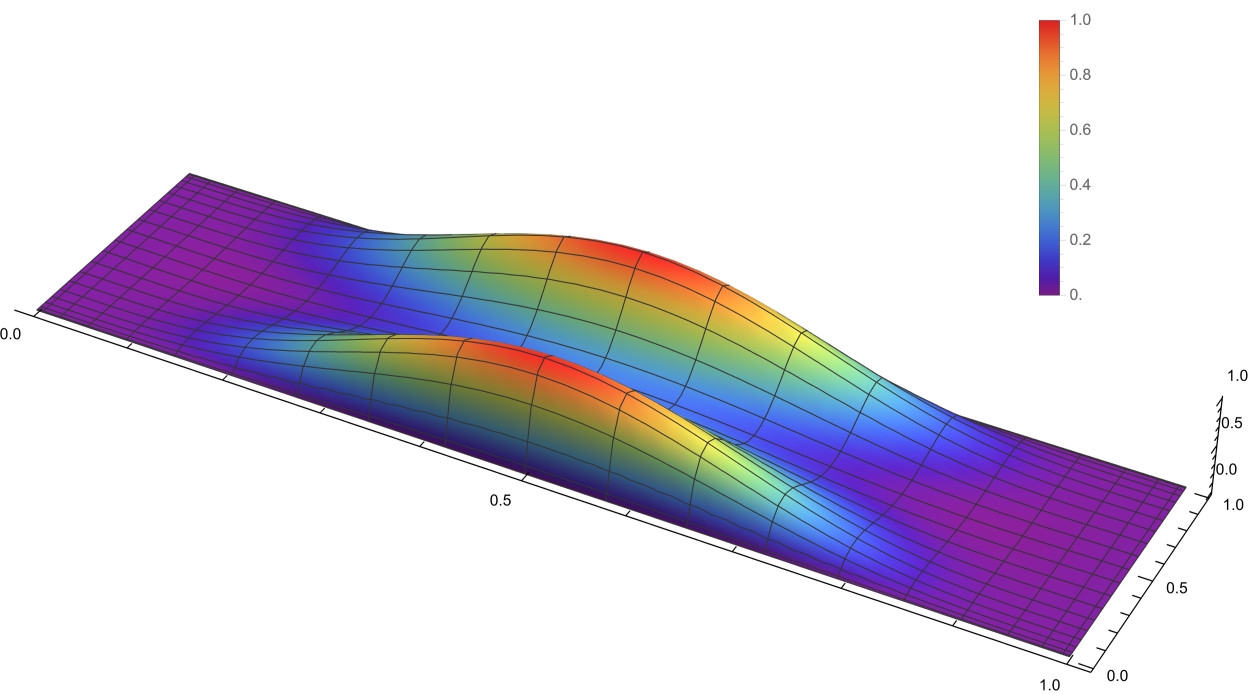}}}\\
		\subfloat[][$\Delta\theta_1$ for CF at $p_a=1.01\times10^{5}$\,Pa.\label{fig9e}]{\scalebox{0.95}{\includegraphics[width=0.5\textwidth]{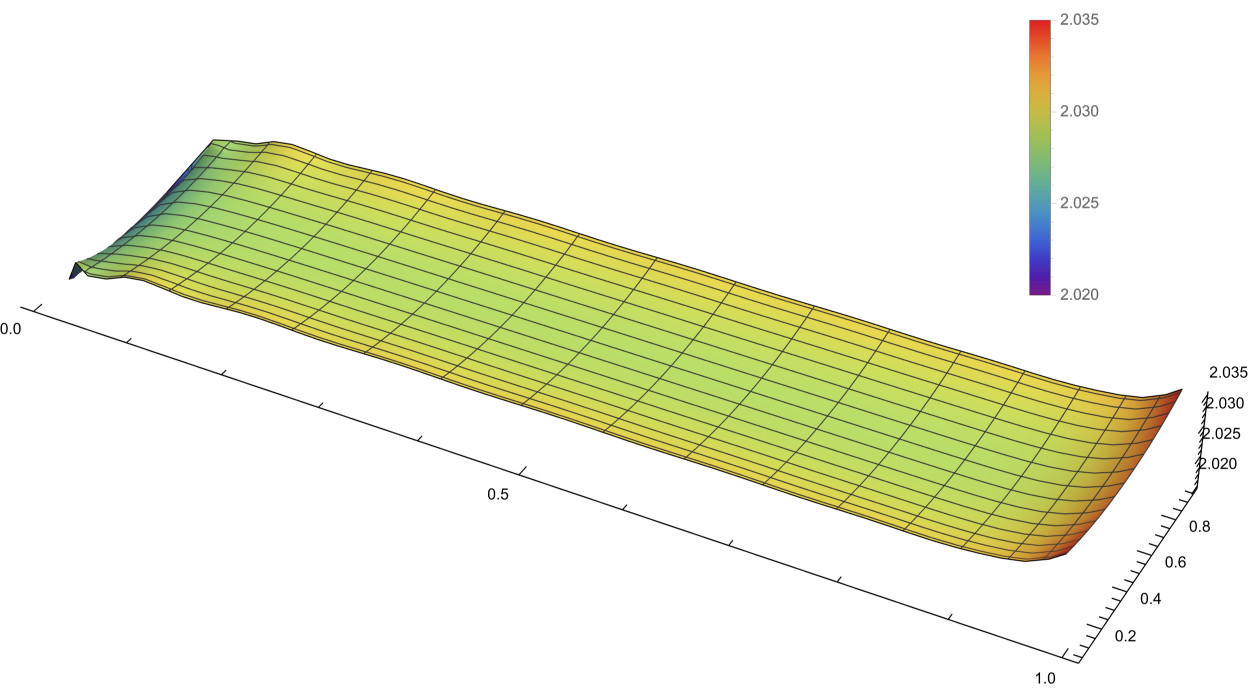}}} \quad
		\subfloat[][$D_1$ for CF at $p_a=1.01\times10^{5}$\,Pa.\label{fig9f}]{\scalebox{0.95}{\includegraphics[width=0.5\textwidth]{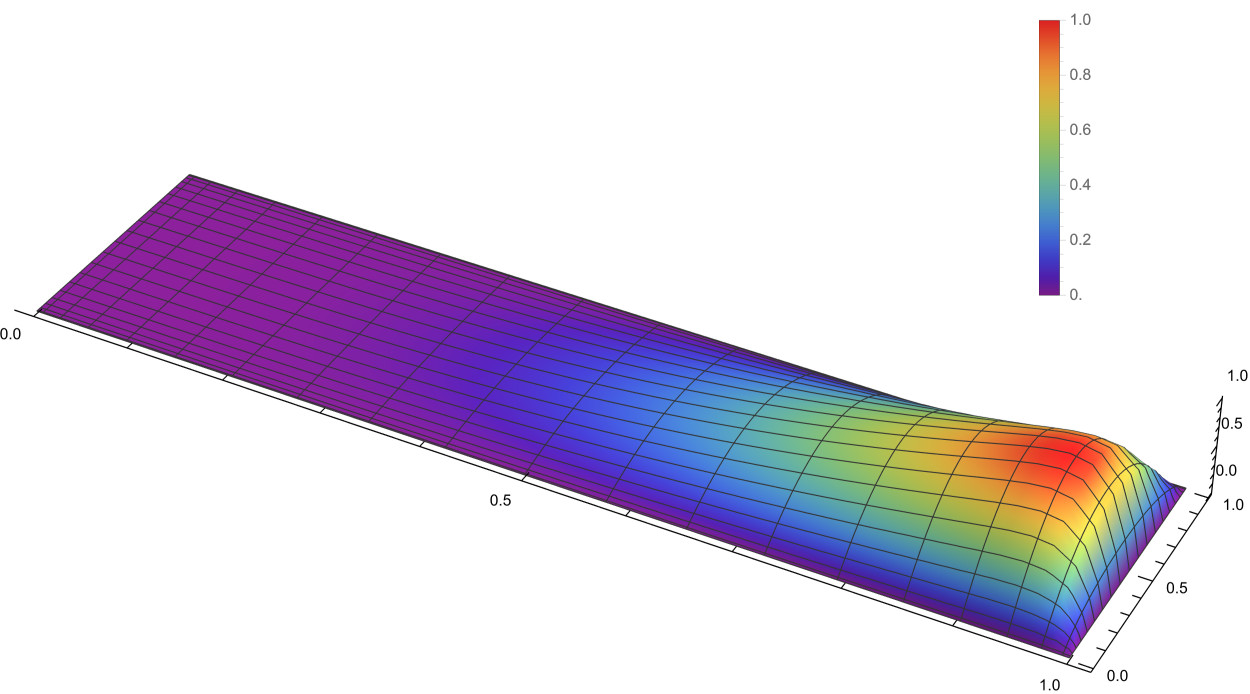}}}
		\caption{Relative phase $\Delta\theta_1$ between the pressure and the deflection (left) and normalized dissipation density $D_1$ (right) of the first coupled mode over the film.}
		\label{fig9}
	\end{figure}

	For the CC microbeam at $p_a=6\times10^{-4}$\,Pa, Figure \ref{fig9a} shows the phase difference distribution of the fundamental mode. $\Delta\theta_1$ lies between $1.15$ and $1.38$\,rad over the film and averages $1.297$\,rad. The film therefore acts predominantly as a damper with a small elastic component, and deviates most from quadrature near the sealed ends, where the air cannot escape along $x$. At atmospheric pressure the phase difference distribution of the fundamental mode, depicted by Figure \ref{fig9c}, changes. In this case, $\Delta\theta_1$ varies by about $2$\,rad across the film, from $-1.32$ to $0.71$\,rad, with an average of $0.10$\,rad, so that the film now acts mainly as a spring, consistently with the increase of the frequency in Figure \ref{fig3}. Within about $0.17$ of each sealed end, $\Delta\theta_1$ becomes negative, and the film locally returns energy to the structure, since the deflection is small near the clamps, these regions contribute little, and the dissipation integrated over the film remains positive. Different regions of the film therefore return the work done on them with different delays, so that the gas layer cannot be represented by a single damping coefficient acting on the microbeam, whatever its value.
	
	\vspace{1cm}
	The phase alone does not indicate where the energy is actually dissipated. For this we estimate the energy exchanged between the microbeam and the film, which came from the work done by the pressure over a cycle. At a given point, the deflection and the pressure of a coupled mode oscillate at the same frequency with different phases. The work done over one cycle is estimated through the area representing the complex amplitudes of the deflection and of the pressure as two vectors in the complex plane, this area is equal to the norm to their cross product. Its integral over the film is proportional to the energy dissipated per cycle by the squeeze film. The local dissipation density is therefore defined as
	\begin{equation} \label{eq67}
		D_r(x,y) = \left| {\bf \Phi}^{\bf 2}_r(x,y) \right|
		\left| {\bf \Phi}^{\bf 1}_r(x) \right|
		\sin \Delta\theta_r(x,y)
		= \mathrm{Im}\left( {\bf \Phi}^{\bf 2}_r(x,y)\,
		\overline{{\bf \Phi}^{\bf 1}_r(x)} \right)
	\end{equation}
	
	Figure \ref{fig9b} shows $D_1$, normalized by its maximum, for the CC microbeam at $p_a=6\times10^{-4}$\,Pa. The dissipation is concentrated at midspan, where the film is most strongly squeezed and furthest from the open edges, and decays towards the clamps. At atmospheric pressure (Figure \ref{fig9d}) the distribution is reversed across the width. The dissipation is concentrated near the open edges and vanishes at the center of the film, where the gas is trapped and acts as a spring rather than as a damper. For the CF microbeam (Figure \ref{fig9f}) the maximum is displaced towards the free end, following the amplitude of the mode, and vanishes at the free end itself, where the pressure is released.
	
	These cases summarize how the boundary conditions of the film redistribute the dissipation along the microbeam. The map of $D_1$ can therefore guide the placement of perforations or venting openings in the microbeam, and are expected to raise the quality factor of the mode, whereas openings in regions where $D_1$ is small or negative mainly alter the elastic action of the film.
	
	%%%%%%%%%%%%%%%%%%%%%%%%%%%%%%%%%%%%%%%%%%%%%%%%%%%%%%%%%%%%%%%%%%%%%%%%%%%%%%%%%%%%%%%%%%%%%%%%%%%%%%%%%%%%%%%%%%%%%%%%%%%%%%%%%
	\section{Reduced-order model based on the complex modes of the coupled system}\label{sec:rom}
	In this section the reduced-order model is derived by rewrite the governing equations in state-space form, consistently with the eigenvalue problem of Section \ref{sec:evp}. The state vector is then expanded on the right eigenvectors of the coupled problem and the residual is projected onto the left eigenvectors, which decouples the linear dynamics exactly and leaves a small set of modal equations driven by the nonlinear terms and the excitation.
	
	\subsection{State-space form of the governing equations}
	For conciseness, the governing equations of the CC and CF microbeams, Equations \eqref{eq1}--\eqref{eq2} and \eqref{eq5}--\eqref{eq6}, are combined into the single form
	\begin{align} 
		&\ddot w + \gamma\dot w + {w_{,xxxx}} = \alpha_{B} w_{,xx}  \left( {{N} - {\alpha _1}\int_0^1 w\,w_{,xx}dx}  \right)  \label{eq68} \\
		& \hspace{2.8cm}  + {\alpha _2}\frac{{{{\left( {{V_{DC}} + V_{AC}(t)} \right)}^2}}}{{{{\left( {1 - w} \right)}^2}}}\left( {1 + \alpha _3 (1 - w) } \right) -{\alpha _4}\int_0^1 {p dy} \nonumber \\
		& \left[ {{(1-w)^3} (1+p)  p_{,x}} \right]_{,x} + \alpha_5 \left[ {{(1-w)^3} (1+p) p_{,y}} \right]_{,y} = \sigma \Big((1-w) \dot{p} - (1+p)\dot{w} \Big) \label{eq69}
	\end{align}
	where $\alpha_{B}=1$ for the CC case and $\alpha_{B}=0$ for the CF case.
	
	The deflection is decomposed into its static and dynamic components given by $w(x,t)=u(x)+q(x,t)$, where $u$ is the static deflection computed at the root-mean-square voltage $V_{RMS}$ given in Section \ref{Subsec_Static}. Using a state-space form the resulting governing equations are given by
	\begin{align}
		&\dot{z}_1 = z_3 \label{eq72} \\
		&\dot{z}_2 = -\frac{3}{\sigma} (u_{,x} +z_{1,x}) (1- u -z_1) (1+z_2) z_{2,x} \label{eq73} \\
		& \hspace{0.8cm} + \frac{1}{\sigma} (1- u -z_1)^2 (z_{2,x}^2 + z_{2,xx} + z_2 z_{2,xx} + \alpha_5 z_{2,y}^2 + \alpha_5 z_{2,yy} + \alpha_5 z_2 z_{2,yy} )   \nonumber\\
		&  \hspace{0.8cm} + \frac{(1+z_2)z_3}{1- u - z_1} \nonumber\\
		&\dot{z}_3 = - \gamma z_3 - {u_{,xxxx}} - {z_{1,xxxx}}  \label{eq74} \\
		& \hspace{0.8cm}+ \alpha_B (u_{,xx} + z_{1,xx} )  \left( {{N} - {\alpha _1} \int_0^1 (u+z_1)(u_{,xx}+z_{1,xx}) \, dx}  \right) \nonumber \\
		& \hspace{0.8cm}  + {\alpha _2}\frac{{{{\left( {{V_{DC}} + V_{AC}(t)} \right)}^2}}}{{{{\left( {1 - u- z_1} \right)}^2}}}\left( {1 + \alpha _3 (1 - u -z_1) } \right) -{\alpha _4}\int_0^1 {z_2 dy} \nonumber
	\end{align}
	where $z_1=q(x,t)$, $z_2=p(x,y,t)$ and $z_3=\dot{q}(x,t)$.
	
	\subsection{Galerkin procedure}
	The state vector is expanded on the right eigenvectors ${\bf \Phi}$ of the condensed matrix ${\bf \hat B}$. Since the governing equations have already been discretized by the differential quadrature method, the expansion and the projection are carried out directly at the DQM grid. The vector ${\bf z}$, collecting the values of $z_1$, $z_2$ and $z_3$ at the grid points used by the right eigenvector, is expanded as
	\begin{equation}\label{eq75}
		{\bf z} = \left( \begin{array}{c} {\bf z^1} \\ {\bf z^2} \\ {\bf z^3} \end{array}\right)
		= \sum_{r \in \mathcal{I}_c} 2\,\mathrm{Re} \left( \eta_{r}(t)\, {\bf \Phi}_r\right)  + \sum_{r \in \mathcal{I}_f} \eta_{r}(t)\, {\bf \Phi}_r
	\end{equation}
	where $\mathcal{I}_c$ and $\mathcal{I}_f$ denote the sets of retained coupled and fluid modes, defined as in Section \ref{sec:modes}. Since complex eigenvectors occur in conjugate pairs, taking twice their real part restores the contribution of the conjugate modes that are not integrated and guarantees that ${\bf z}$ is real at all times. Modes with real eigenvalues are retained once and contribute directly. The coordinates $\eta_r$ are therefore complex for $r\in\mathcal{I}_c$ and real for $r\in\mathcal{I}_f$, and the reduced system has $N_R=2n_C+n_F$ real unknowns corresponding to the number of retained modes, with $n_C \le \mathrm{card}(\mathcal{I}_c)$ and $n_F \le \mathrm{card}(\mathcal{I}_f)$.
	
	Equation \eqref{eq75} is substituted into Equations \eqref{eq72}--\eqref{eq74}, and the residual is premultiplied by the transpose of the left eigenvectors ${\bf \Psi}_s$. Writing the right-hand side of Equations \eqref{eq72}--\eqref{eq74} as ${\bf \hat F}({\bf z},t) = {\bf \hat B}{\bf z} + {\bf \hat N}({\bf z},t)$, the projection gives
	\begin{equation}\label{eq76}
		{\bf \Psi}^T_s \dot{\bf z} = {\bf \Psi}^T_s {\bf\hat B} {\bf z} + {\bf \Psi}^T_s {\bf\hat N}\left( {\bf z} \right), \quad s \in \mathcal{I}_c \cup \mathcal{I}_f
	\end{equation}
	where ${\bf \hat N}={\bf \hat F}-{\bf \hat B}{\bf z}$ collects the terms of the right-hand side that are not contained in the linear operator. Its first partition vanishes, since Equation \eqref{eq72} is linear:
	$${\bf\hat N}= \left( \begin{array}{c} 
		{\bf 0} {}^{(n_w \times 1)} \\ 
		{\bf\hat N^2} {}^{(n_p \times 1)} \\ 
		{\bf\hat N^3} {}^{(n_w \times 1)} 
	\end{array}  \right). $$
	Projecting onto ${\bf \Psi}_s$ and applying the biorthonormality conditions \eqref{eq54} and \eqref{eq55} eliminates not only the contributions of the other retained modes but also those of the conjugate modes. The projection in Equation \eqref{eq76} becomes
	\begin{equation}\label{eq77}
		\dot{\eta}_s = \lambda_s {\eta}_s + {{\bf \Psi}^{\bf{2}}_s}^T {\bf\hat N^2}\left( {\bf z} \right)  + {{\bf \Psi}^{\bf{3}}_s}^T {\bf\hat N^3}\left( {\bf z} \right), \quad s \in \mathcal{I}_c \cup \mathcal{I}_f
	\end{equation}
	
	The right-hand side ${\bf \hat F}$ is evaluated at the grid points and condensed as in Equation \eqref{eq58}, the pressure equation \eqref{eq73} being already in explicit form. Subtracting ${\bf \hat B}{\bf z}$ ensures that the linear dynamics about the static deflection is carried exactly by the eigenvalues $\lambda_j$. Since ${\bf \hat F}$ contains the full time-dependent voltage, whereas ${\bf \hat B}$ results from a linearization about the static deflection under $V_{RMS}$, ${\bf \hat N}$ does not vanish for ${\bf z}={\bf 0}$. The two partitions of ${\bf \hat N}$ are given in vector form by
	\begin{align}
		{\bf\hat N^2} \left( {\bf z} \right) = &-\frac{3}{\sigma} {\bf E}  \left( {\bf a}^{(1)} {\bf u} + {\bf a}^{(1)} {\bf z^1} \right) \circ \left( {\bf EH} \right)   \circ ( {\bf 1} + {\bf z^2} ) \circ \left( {\bf b}_x^{(1)} {\bf z^2}\right)  \label{eq78} \\
		& + \frac{1}{\sigma} \Big[ \left( {\bf EH} \right)^{\circ 2} \circ \Big( \left( {\bf b}_x^{(1)} {\bf z^2} \right)  ^{\circ 2} + {\bf b}_x^{(2)} {\bf z^2} + {\bf z^2} \circ \left( {\bf b}_x^{(2)} {\bf z^2} \right) \nonumber \\
		&  + \alpha_5 \left( {\bf b}_y^{(1)} {\bf z^2} \right) ^{\circ 2} + \alpha_5 {\bf b}_y^{(2)} {\bf z^2} + \alpha_5 {\bf z^2} \circ \left( {\bf b}_y^{(2)} {\bf z^2} \right) \Big)   \Big] \nonumber\\
		& + ({\bf 1} + {\bf z^2}) \circ \left( {\bf E }{\bf z^3}\right)  \circ \left( {\bf EH} \right)^{\circ -1} + {\bf C}_{22}^{-1} {\bf K}_{22} {\bf z}^{\bf 2} + {\bf C}_{22}^{-1} {\bf C}_{21} {\bf z}^{\bf 3} \nonumber \\
		{\bf\hat N^3} \left( {\bf z} \right) = & - \gamma {\bf z^3} - {\bf a}^{(4)} {\bf u} - {\bf a}^{(4)} {\bf z^1}  \label{eq79} \\
		& + \alpha_B ( {\bf a}^{(2)} {\bf u} + {\bf a}^{(2)} {\bf z^1} )  \left( {N} - {\alpha _1} {\bf \tilde T}^T \left( ({\bf u} + {\bf z^1}) \circ ( {\bf a}^{(2)} {\bf u} +{\bf a}^{(2)} {\bf z^1} ) \right) \right) \nonumber \\
		& + {\alpha _2} \left( {{V_{DC}} + V_{AC}(t)} \right)^2 \left( {\bf H} \right)^{\circ -2} \circ \left( {\bf 1} + \alpha_3 {\bf H} \right)  - {\alpha _4} {\bf G} {\bf z^2}  \nonumber \\
		& + {\bf K}_{11} {\bf z}^{\bf 1} + {\bf K}_{12} {\bf z}^{\bf 2} + {\bf C}_{11} {\bf z}^{\bf 3} \nonumber
	\end{align}
	where $\circ$ denotes the elementwise (Hadamard) product of two vectors of the same length and ${\bf v}^{\circ k}$ the elementwise $k^{th}$ power of a vector ${\bf v}$, ${\bf H} = {\bf 1}- {\bf u} - {\bf z^1}$ is the gap, and ${\bf 1}$ is a vector of ones of the appropriate length. The operators ${\bf E}$ and ${\bf G}$ transfer quantities between the nodes of the microbeam and those of the film, and ${\bf a}^{(k)}$, ${\bf b}_x^{(k)}$ and ${\bf b}_y^{(k)}$ are the condensed derivative matrices, all are defined in \ref{ApII}. Since the deflection vanishes at both clamped ends, the stretching integral involves the interior quadrature weights only, $\tilde{\bf T} = (T_2, \dots, T_{n-1})^T$, with $T_i$ given by Equation \eqref{eqNC}.
	
	Substituting Equations \eqref{eq78} and \eqref{eq79} into Equation \eqref{eq77} yields a set of $N_R$ first-order differential equations in the modal coordinates $\eta_i$, driven by the applied voltage $V_{DC} + v_{ac} \cos (\Omega t)$, where $\Omega$ denotes the nondimensional excitation frequency.
	
	The reduced model is integrated in time as a set of nonlinear ordinary differential equations. The number and the nature of the retained modes are not arbitrary. This question, together with the validation of the reduced model against the full discretized system and its computational cost, is addressed in the following section.
	
	%%%%%%%%%%%%%%%%%%%%%%%%%%%%%%%%%%%%%%%%%%%%%%%%%%%%%%%%%%%%%%%%%%%%%%%%%%%%%%%%%%%%%%%%%%%%%%%%%%%%%%%%%%%%%%%%%%%%%%%%%%%%%
	\section{Dynamic response of the reduced-order model}\label{sec:dynamics}
	This section assesses the reduced-order model of Section \ref{sec:rom} in the forced nonlinear regime. The influence of the composition and the size of the retained basis is examined first, by comparing the results with direct integration of the full DQM discretized system.
	
	The microbeams are tested for $n=21$ and $p_a=1.01325\times10^{5}$\,Pa, with $V_{DC}=10$\,V, $v_{ac}=10$\,V and $\Omega=4$ for the CC microbeam, and $V_{DC}=5$\,V, $v_{ac}=5$\,V and $\Omega=2$ for the CF microbeam. The static solution about which the eigenvalue problem is linearized is computed at the root-mean-square voltage $V_{RMS}=\sqrt{V_{DC}^2 + \tfrac12 v_{ac}^2}$. The natural frequencies and quality factors at this operating point are listed in Table \ref{tab6}.
	
	The accuracy of the reduced model depends on the composition of the retained basis and not only on its size. The convergence analysis is reported in Table \ref{tab7}, where $n_C$ and $n_F$ denote the number of retained coupled and fluid modes respectively. The reference solution is obtained by direct time integration of the full DQM discretized system, Equations \eqref{eq12}--\eqref{eq13} for the CC microbeam and \eqref{eq16}--\eqref{eq17} for the CF microbeam, whose dimension is $2n_w+n_p=228$ first-order states for $n=21$.
	
	\begin{table}[H]
		\centering
		\caption{Natural frequencies and quality factors of the modes linearized about the static deflection obtained at the effective voltage $V_{RMS}=\sqrt{V_{DC}^2+\frac12 v_{ac}^2}$, for $n=21$ and $p_a=1.01325\times10^{5}$\,Pa. Only the symmetric modes of the CC microbeam are listed.} \label{tab6}
		\begin{tabular}{ccccc}
			\hline
			Mode  & Nature & Parity & $\omega_i$ &                    $Q_i$                     \\ \hline
			\multicolumn{5}{c}{CC microbeam, $V_{DC}=10$\,V, $v_{ac}=10$\,V and $u_{11}=0.201$} \\
			1   &   F    &   S    &    0.44    &                     0.50                     \\
			3   &   F    &   S    &   13.60    &                     0.50                     \\
			5   &   F    &   S    &   34.38    &                     0.50                     \\
			11-12 &   C    &   S    &   135.39   &                     5.40                     \\
			21-22 &   C    &   S    &   181.57   &                     8.19                     \\
			33-34 &   C    &   S    &   332.75   &                    27.87                     \\ \hline
			\multicolumn{5}{c}{CF microbeam, $V_{DC}=5$\,V, $v_{ac}=5$\,V and $u_{21}=0.178$}  \\
			1-2  &   C    &   --   &    3.03    &                     0.65                     \\
			3-4  &   C    &   --   &   22.14    &                     5.32                     \\
			5-6  &   C    &   --   &   62.10    &                    16.29                     \\
			11   &   F    &   --   &   227.64   &                     0.50                     \\
			12   &   F    &   --   &   248.59   &                     0.50                     \\
			13   &   F    &   --   &   265.91   &                     0.50                     \\ \hline
		\end{tabular}
	\end{table}
	
	\begin{table}[H]
		\centering
		\caption{Convergence of the reduced-order model with the composition of the retained basis, and associated computational cost. Results are obtained for $n=21$ and $p_a=1.01325\times10^{5}$\,Pa, with $V_{DC}=10$\,V, $v_{ac}=10$\,V and $\Omega=4$ for the CC microbeam, and $V_{DC}=5$\,V, $v_{ac}=5$\,V and $\Omega=2$ for the CF microbeam. $n_C$ and $n_F$ denote the number of retained coupled and fluid modes. The mode shapes used for the CC microbeam are all symmetric. Relative and $L^2$ errors are evaluated against the full model at $n=21$, whose dimension is $2n_w+n_p=228$.}
		\label{tab7}
		\begin{tabular}{ccccccccc}
			\hline
			$n_C$ & $n_F$ &         $N_{R}$         & $w_{Max}$ & $\varepsilon_w$ & $\varepsilon_{L^2}$ & $\int p\,dA$ & $\varepsilon_{\int p}$ & CPU  \\
			&       &                         &           &      (\%)       &        (\%)         & $\times10^3$ &          (\%)          & (s)  \\ \hline
			\multicolumn{9}{c}{CC microbeam}                                                      \\
			1   &   0   &            2            & 0.214467  &    $-13.95$     &        14.96        &   8.15593    &        $-0.23$         & 1.9  \\
			1   &   1   &            3            & 0.247320  &     $-0.76$     &        1.73         &   8.15794    &        $-0.20$         & 3.3  \\
			3   &   1   &            7            & 0.247541  &     $-0.68$     &        1.34         &   8.50282    &        $+4.02$         & 119  \\
			3   &   2   &            8            & 0.246105  &     $-1.25$     &        1.54         &   8.08549    &        $-1.09$         & 226  \\
			3   &   3   &            9            & 0.245832  &     $-1.36$     &        1.56         &   8.37111    &        $+2.41$         & 404  \\
			\multicolumn{3}{c}{Full model ($n=9$)}  & 0.249194  &     $-0.01$     &         --          &   8.16936    &        $-0.06$         & 6.8  \\
			\multicolumn{3}{c}{Full model ($n=13$)} & 0.249224  &     $-0.00$     &         --          &   8.17447    &        $-0.00$         &  87  \\
			\multicolumn{3}{c}{Full model ($n=17$)} & 0.249221  &     $-0.00$     &         --          &   8.17449    &        $+0.00$         & 548  \\
			\multicolumn{3}{c}{Full model ($n=21$)} & 0.249226  &       --        &         --          &   8.17448    &           --           & 3227 \\ \hline
			\multicolumn{9}{c}{CF microbeam}                                                      \\
			1   &   0   &            2            & 0.580152  &    $+17.20$     &        13.72        &   0.91283    &        $-22.05$        & 2.3  \\
			1   &   1   &            3            & 0.502130  &     $+1.44$     &        1.57         &   1.12260    &        $-4.14$         & 4.2  \\
			2   &   1   &            5            & 0.502504  &     $+1.51$     &        1.23         &   1.12427    &        $-4.00$         &  24  \\
			3   &   1   &            7            & 0.502626  &     $+1.54$     &        1.26         &   1.12535    &        $-3.90$         & 125  \\
			3   &   2   &            8            & 0.502436  &     $+1.50$     &        1.24         &   1.11954    &        $-4.40$         & 241  \\
			3   &   3   &            9            & 0.499844  &     $+0.98$     &        0.84         &   1.16575    &        $-0.45$         & 287  \\
			\multicolumn{3}{c}{Full model ($n=9$)}  & 0.484333  &     $-2.16$     &         --          &   1.12814    &        $-3.67$         & 8.5  \\
			\multicolumn{3}{c}{Full model ($n=13$)} & 0.493165  &     $-0.37$     &         --          &   1.16398    &        $-0.60$         & 108  \\
			\multicolumn{3}{c}{Full model ($n=17$)} & 0.494829  &     $-0.04$     &         --          &   1.17040    &        $-0.06$         & 561  \\
			\multicolumn{3}{c}{Full model ($n=21$)} & 0.495009  &       --        &         --          &   1.17106    &           --           & 3650 \\ \hline
		\end{tabular}
	\end{table}
	
	\vspace{1cm}
	Three quantities are evaluated over the last three periods of the steady-state response: (i) the maximum deflection $w_{Max}$, taken at midspan for the CC microbeam ($w_{11}$) and at the free end for the CF microbeam ($w_{21}$), (ii) the relative error and $L^2$ deviation of the deflection history from the reference ($n=21$),
	\begin{equation}\label{eq81}
		\varepsilon_{w} = \frac{w_R - w_F}{w_F} \quad \mbox{and} \quad \varepsilon_{L^2} = \left( \frac{\int (w_R - w_F)^2\,dt}{\int w_F^2\,dt} \right)^{1/2}
	\end{equation}
	where $w_R$ and $w_F$ are the maximum deflections of the reduced and full models, respectively, (iii) and the maximum of the resultant force exerted by the film on the microbeam, $\int p\,dx\,dy$. 
	
	CPU times are obtained on the same machine with identical integrator settings and tolerances. Since the electrostatic load is uniform along the microbeam, it does not excite the antisymmetric modes of the CC microbeam, whose basis is therefore restricted to symmetric modes.
	
	Because the projection is of Petrov--Galerkin type and the underlying operator is non-self-adjoint, the projection minimizes only the component of the residual lying in the test space, and not the residual itself, so that no variational principle bounds the error of the reduced model \citep{aresdeparga2023hyper,grimberg2020stability}. In the present case the convergence with respect to the number of retained modes is found to be non-monotonic, as Table \ref{tab7} shows.
	
	For the CC microbeam the deviation on $\int p\,dA$ has successive alternating sign. A single configuration exhibiting a small error is therefore not sufficient evidence of convergence, since it may result from a fortuitous cancellation. What must be assessed is the trend over a sequence of bases, and in particular the amplitude of the oscillation rather than the value of any individual entry.
	
	The first line of each block of Table \ref{tab7} establishes the central result of this work. A basis composed exclusively of coupled modes fails to reproduce the response. In fact, the deviation on the deflection reaches $-13.95$\% for the CC microbeam and $17.20$\% for the CF microbeam. It is not reduced by the addition of further coupled modes. Adding a single fluid relaxation mode, on the other hand, brings the deviation below $1.5$\% in both configurations, with a reduced model of only three real degrees of freedom against $228$ for the full system.
	
	The two microbeams fail in opposite directions when the fluid modes are dropped. The amplitude of the response is underestimated for the CC microbeam and overestimated for the CF microbeam. For the CC microbeam, the first fluid mode relaxation time exceeds the excitation period (Table \ref{tab6}), so that the film cannot drain within a cycle. Therefore, if the first coupled mode is accounted alone in the basis, it embodies the stiffness of the film at its own frequency and the gas is effectively trapped. While integrating the fluid mode part of the gas escapes and the stiffness of the film is not overestimated.
	
	For the CF microbeam the situation is reversed. Its fluid modes lie two orders of magnitude above the excitation frequency and their relaxation time is negligible compared with the period of the excitation. The contribution of the fluid modes is mainly coming from their projection on the forcing. They act as quasi-static corrections to the pressure distribution, which the coupled mode cannot represent by itself. In their absence the resultant film force is underestimated, and the deflection is overestimated accordingly.
	
	Beyond the first fluid mode the two configurations behave differently. For the CC microbeam the deviation on the maximum deflection remains between $0.68$ and $1.36$\% for all richer bases. Further fluid modes slightly degrade the result rather than improve it. For the CF microbeam, enriching the basis with coupled modes or with a second fluid mode leaves the deviation close to $1.5$\% and that on the resultant force close to $4$\%, whereas the third fluid mode reduces them to $0.98$ and $0.45$\% respectively. 
	
	The $L^2$ deviations reported in Table \ref{tab7} confirm these conclusions. For the bases without fluid modes they reach $14.96$ and $13.72$\% for the CC and CF microbeams, of the same order as the deviations on the peak values, which shows that the failure affects the whole waveform and is not an artifact of extracting a single extremum. A single fluid mode reduces them by a factor of about nine in both cases, to $1.73$ and $1.57$\%.
	
	Concerning the computational times, the DQM discretization is itself spectrally efficient, so that the full model already provides an accurate solution at $n=9$ in a few seconds. The advantage of the reduced-order model lies in the analyses for which a low-dimensional model is a prerequisite, such as frequency sweeps, continuation, bifurcation tracking and extensive parametric studies.
	
\begin{figure}[H]	
		\centering
		\subfloat[][CC microbeam at midspan, for $V_{DC}=10$\,V, $v_{ac}=10$\,V, $\Omega=4$.\label{fig10a}]
		{\scalebox{0.95}{\includegraphics[width=0.5\textwidth]{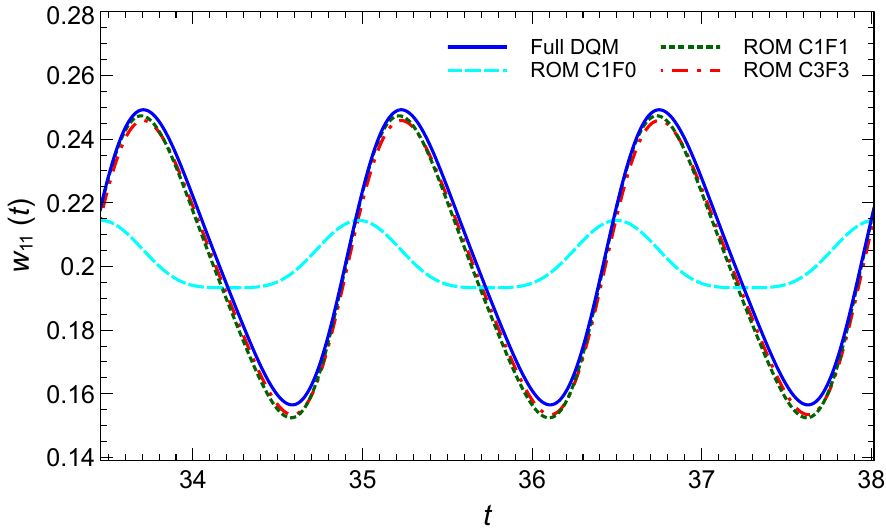}}}
		\quad
		\subfloat[][CF microbeam at free end, for $V_{DC}=5$\,V, $v_{ac}=5$\,V, $\Omega=2$.\label{fig10b}]
		{\scalebox{0.95}{\includegraphics[width=0.5\textwidth]{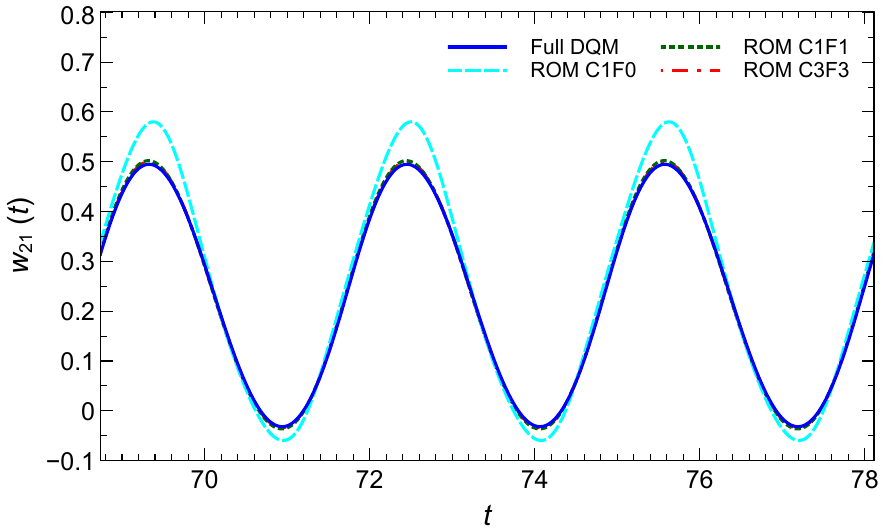}}}
		\caption{Steady-state deflection over the last three excitation periods. C$n_C$F$n_F$ denotes a reduced basis containing $n_C$ coupled and $n_F$ fluid modes. $p_a=1.01325\times10^{5}$\,Pa and $n=21$.}
		\label{fig10}
	\end{figure}

	Figure \ref{fig10} shows the steady-state response over the last three periods for three representative bases. The reduced models containing at least one fluid mode are nearly indistinguishable from the reference, in amplitude and in phase. The basis restricted to coupled modes is different. For the CC microbeam its amplitude is about four times smaller than the reference with a distorted waveform. For the CF microbeam it overestimates the maxima by $17$\%.
	
\begin{figure}[H]	
		\centering
		\subfloat[][CC microbeam, full DQM model.\label{fig11a}]
		{\scalebox{0.95}{\includegraphics[width=0.5\textwidth]{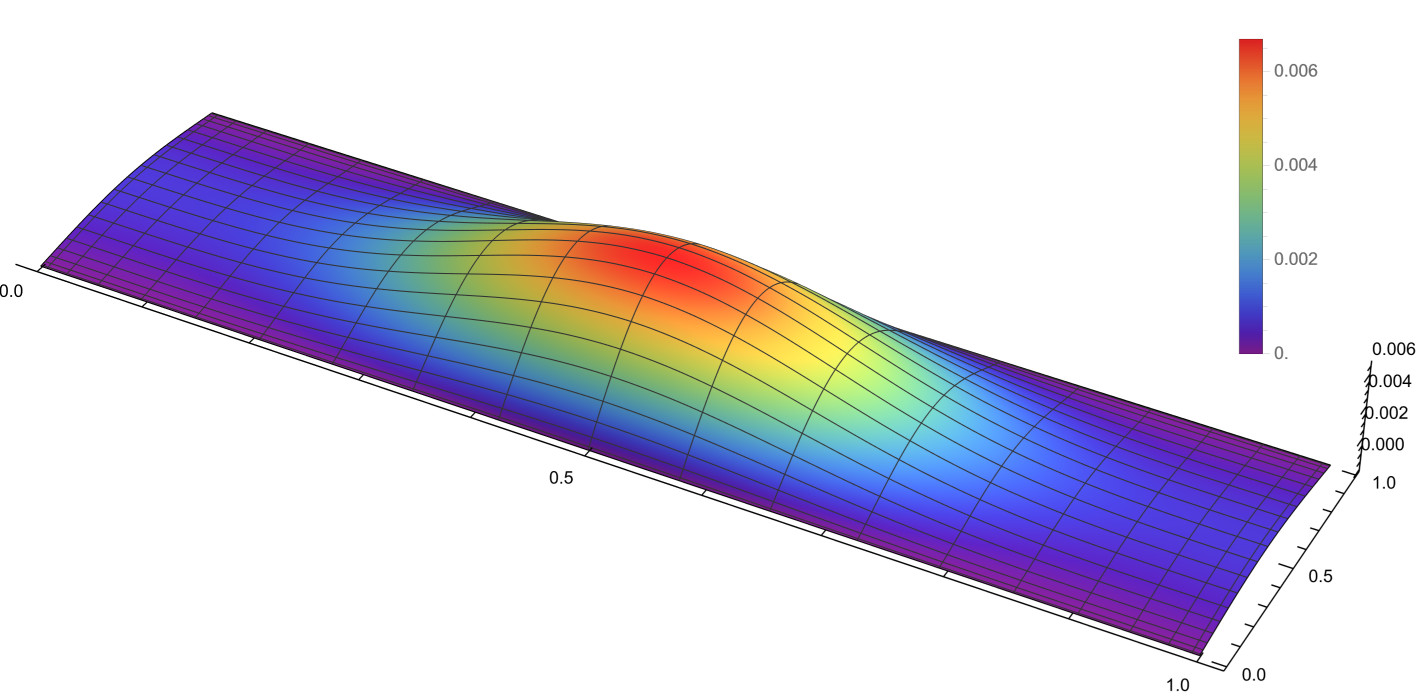}}}
		\quad
		\subfloat[][CF microbeam, full DQM model.\label{fig11b}]
		{\scalebox{0.95}{\includegraphics[width=0.5\textwidth]{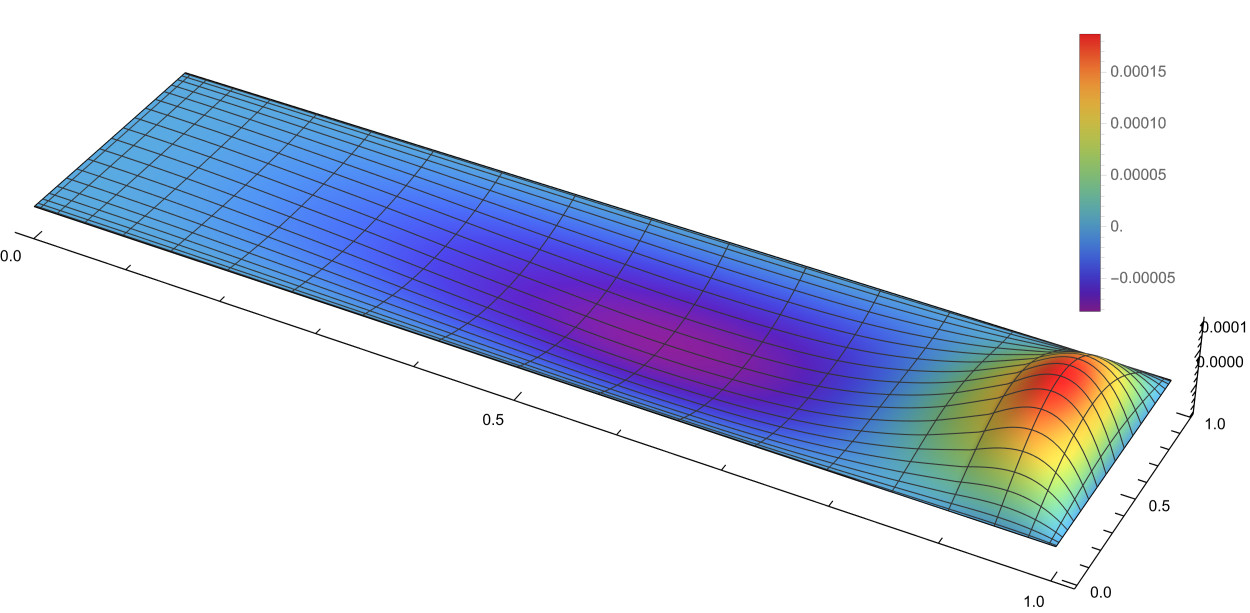}}}\\
		\subfloat[][CC microbeam, C1F1.\label{fig11c}]
		{\scalebox{0.95}{\includegraphics[width=0.5\textwidth]{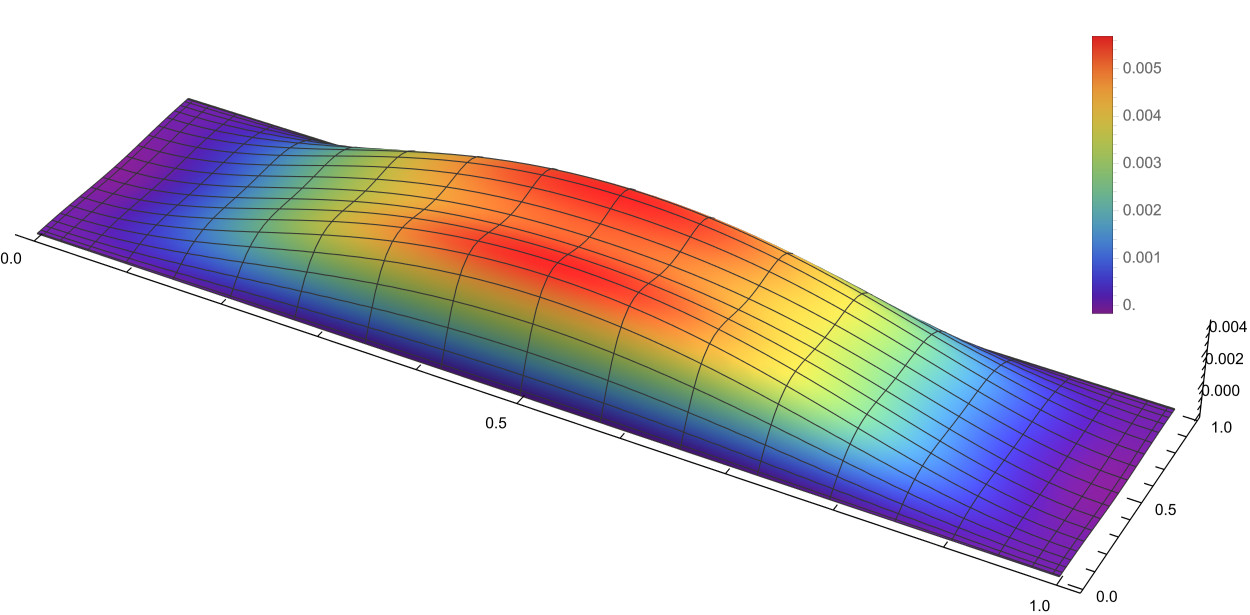}}}
		\quad
		\subfloat[][CF microbeam, C1F1.\label{fig11d}]
		{\scalebox{0.95}{\includegraphics[width=0.5\textwidth]{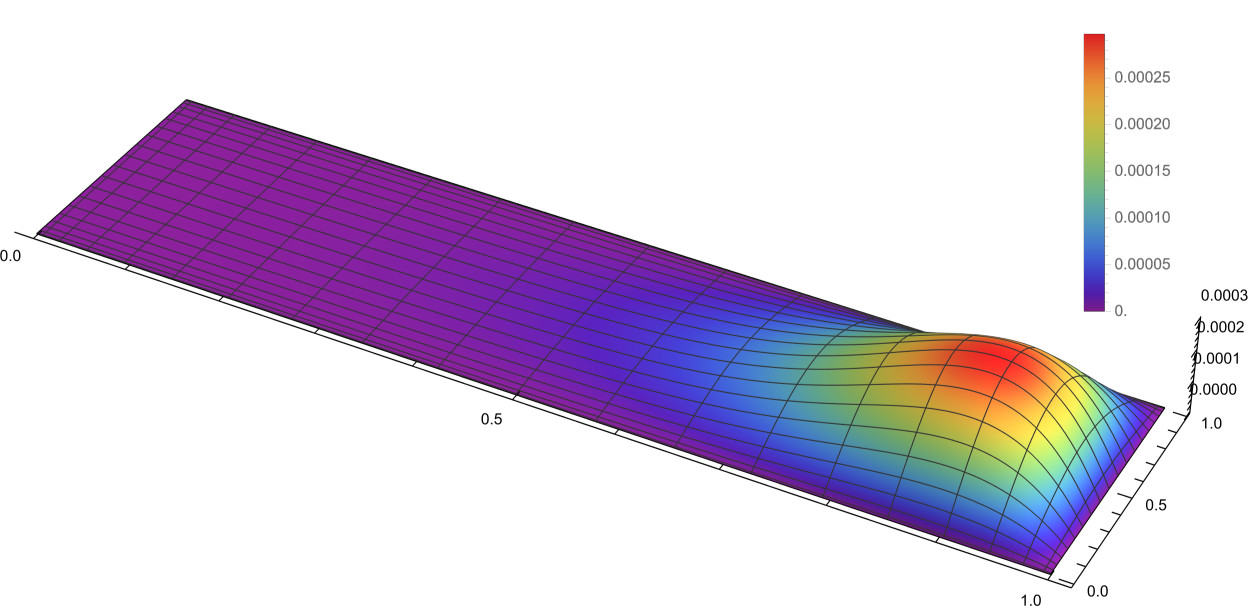}}}\\
		\subfloat[][CC microbeam, C3F3.\label{fig11e}]
		{\scalebox{0.95}{\includegraphics[width=0.5\textwidth]{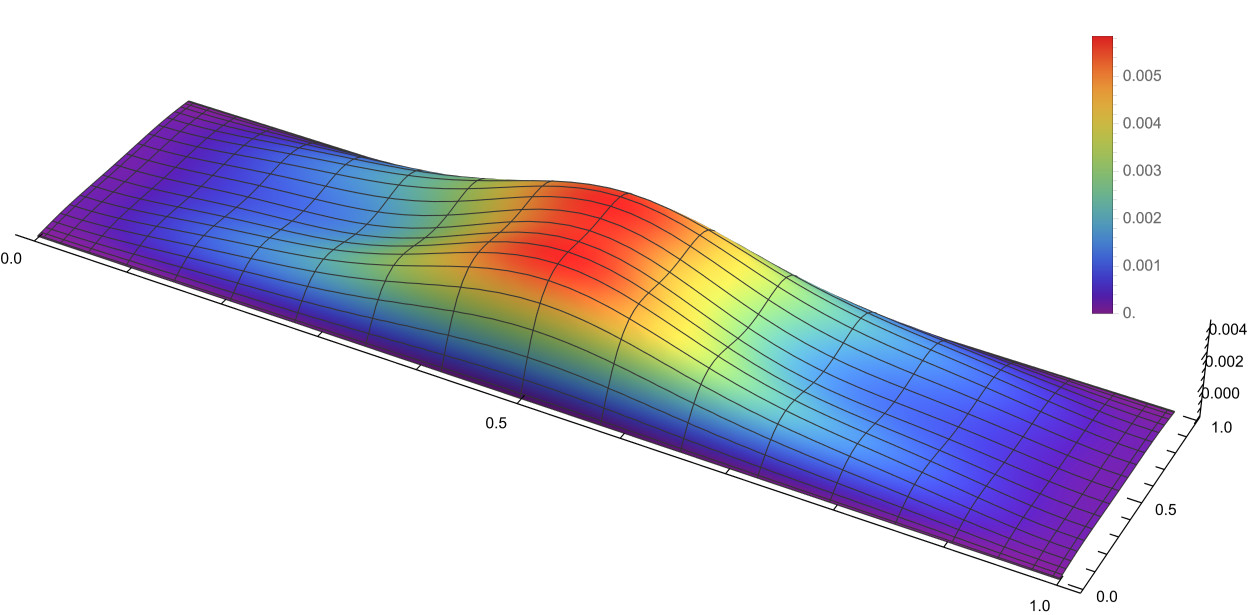}}}
		\quad
		\subfloat[][CF microbeam, C3F3.\label{fig11f}]
		{\scalebox{0.95}{\includegraphics[width=0.5\textwidth]{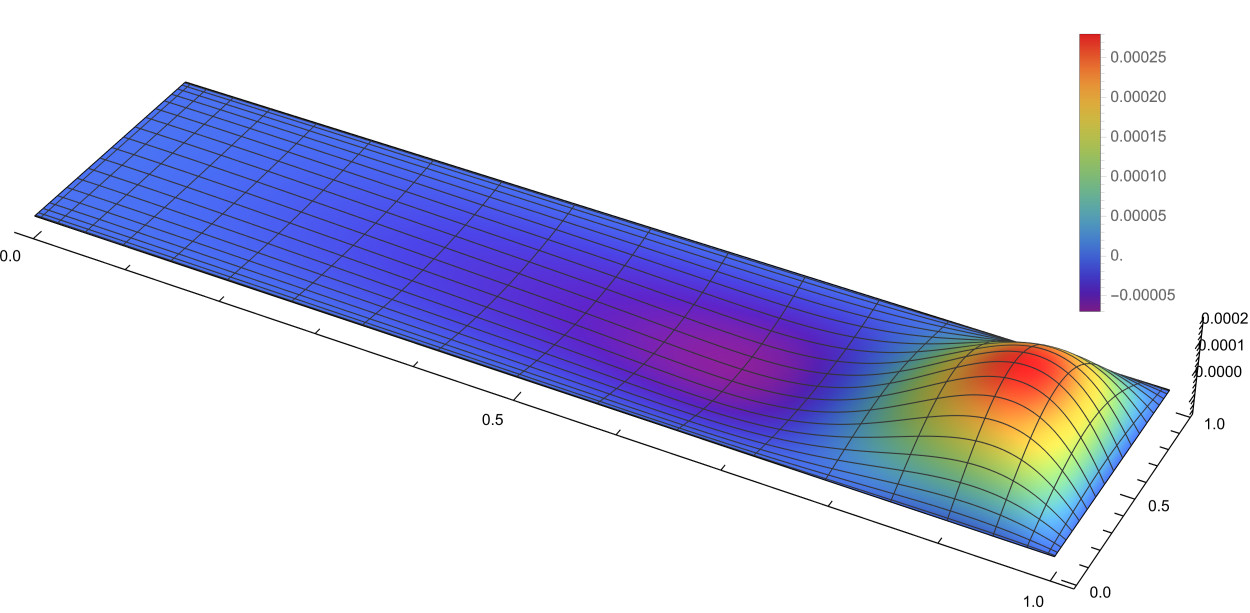}}}
		\caption{Relative pressure distribution over the film at the instant of maximum deflection of each model, for $n=21$ and $p_a=1.01325\times10^{5}$\,Pa, with $V_{DC}=10$\,V, $v_{ac}=10$\,V, $\Omega=4$ for the CC microbeam and $V_{DC}=5$\,V, $v_{ac}=5$\,V, $\Omega=2$ for the CF microbeam. C$n_C$F$n_F$ denotes a reduced basis containing $n_C$ coupled and $n_F$ fluid modes.}
		\label{fig11}
	\end{figure}

	Figure \ref{fig11} compares the pressure fields, each shown at the instant of maximum deflection of the corresponding model. For the CC microbeam the reference field is concentrated around midspan and decays rapidly towards the clamped ends. With $n_C=n_F=1$ the reduced model distributes the pressure more broadly, with a lower maximum and higher values towards the clamps, since a single fluid mode cannot localize the field, with $n_C=n_F=3$ the localization is recovered. For the CF microbeam the reference field is positive near the free end, where the gap is smallest, and negative over the central part of the film. The basis with a single fluid mode reproduces the positive region but not the depression, which is recovered with three fluid modes. The spatial distribution of the pressure therefore requires a richer set of fluid modes than the deflection, consistently with the fact that the microbeam responds to the integral of the pressure over the film rather than to its local value.
	
	%%%%%%%%%%%%%%%%%%%%%%%%%%%%%%%%%%%%%%%%%%%%%%%%%%%%%%%%%%%%%%%%%%%%%%%%%%%%%%%%%%%%%%%%%%%%%%%%%%%%%%%%%%%%%%%%%%%%%%%%%%%%%
	\section{Conclusions}\label{sec:conclusions}
	
	A reduced-order model has been propsoed for electrostatically actuated microbeams subjected to squeeze-film damping, in which the reduced basis is formed by complex modes of the fully coupled Reynolds fluid--structure problem. The coupled system is first discretized by the differential quadrature method (DQM) for the static and eigenvalue problem (EVP). Because the microbeam is loaded by the pressure while the air film is driven by the microbeam velocity, the resulting EVP operator is non-self-adjoint. Its right and left eigenvectors are made biorthonormal and then used in a Petrov--Galerkin projection to generate a reduced-order model of the coupled problem. The EVP generated two sets of modes with distinct nature. A complex-conjugate pairs that described coupled oscillations of the microbeam and the air film, and real modes describing non-oscillatory relaxations of the pressure field. The frequencies and quality factors of the coupled modes have been compared with the experimental and numerical results available in the literature.
	
	The analysis of the modal assurance criterion and of the relative phase between the modes demonstrated a non-classical aspect of the damping introduced by the fluid modes. At atmospheric pressure the structural coupled modes remain nearly real, the microbeam modes are almost in-phase with each others. Their pressure fields, by contrast, admit no real representation. At low pressure the film acts almost as a pure damper, so that the high quality factor observed there results from the small amplitude of the film reaction rather than from a small phase difference. The local dissipation density showed that, for the clamped--clamped microbeam, the energy is lost at midspan at low pressure and near the open edges at atmospheric pressure, whereas for the clamped--free microbeam it is displaced towards the free end. No single lumped damping coefficient can represent such a film.
	
	The dynamic analysis at atmospheric pressure established the central result of this work. A basis composed exclusively of coupled modes fails to reproduce the forced nonlinear response, the deviation on the maximum deflection reaching $14$ and $17$\% for the two microbeams, and it is not reduced by adding further coupled modes. The inclusion of a single fluid relaxation mode brings the deviation below $1.5$\%, with a reduced model of three real degrees of freedom against $228$ first-order states for the full discretized system.
	
	At equal resolution the reduced model with one coupled and one fluid mode is integrated about a thousand times faster than the full system, and its basis, computed once for a given operating point, serves all the responses at that point. Since the DQM is itself spectrally efficient, the advantage of the reduction lies less in the integration of a single response than in the analyses for which a low-dimensional model is a prerequisite. Future work will therefore address the use of the reduced model for frequency sweeps, continuation and bifurcation analysis near pull-in, including forced responses near the structural resonance and within the relaxation band of the film.
	
	%%%%%%%%%%%%%%%%%%%%%%%%%%%%%%%%%%%%%%%%%%%%%%%%%%%%%%%%%%%%%%%%%%%%%%%%%%%%%%%%%%%%%%%%%%%%%%%%%%%%%%%%%%%%%%%%%%%%%%%%%%%%%
	\appendix
	
	\section{DQM and LaDQM matrices} \label{ApI}
	
	\begin{align}
		& A_{ij}^{(1)} =  {\frac{{\prod\limits_{v = {0};v \ne i}^{{n+1}} {({\xi _i} - {\xi _v})} }}{{({\xi _i} - {\xi _j})\prod\limits_{v = {0};v \ne j}^{{n+1}} {({\xi _j} - {\xi _v})} }}}  && {0} \le j \le {n+1}, \quad j \ne i \label{Ap01} \\
		& A_{ij}^{(k)} = k\left( A_{ii}^{(k-1)}A_{ij}^{(1)} - \frac{A_{ij}^{(k-1)}}{\xi_i - \xi_j} \right) && 0 \le j \le n+1, \quad j \ne i \\
		& A_{ii}^{(k)} = -\sum\limits_{v=0;\,v\ne i}^{n+1} A_{iv}^{(k)} && 0 \le i \le n+1 \\
		&	&& k = 2,3, \ldots \nonumber
	\end{align}
	
	\begin{align}
		& B_{ij}^{(1)} =  {\frac{{\prod\limits_{v = {1};v \ne i}^{{n}} {({\xi _i} - {\xi _v})} }}{{({\xi _i} - {\xi _j})\prod\limits_{v = {1};v \ne j}^{{n}} {({\xi _j} - {\xi _v})} }}}  && {1} \le j \le {n}, \quad j \ne i \label{Ap02} \\
		& B_{ij}^{(k)} = k\left( B_{ii}^{(k-1)}B_{ij}^{(1)} - \frac{B_{ij}^{(k-1)}}{\xi_i - \xi_j} \right) && 1 \le j \le n, \quad j \ne i \\
		& B_{ii}^{(k)} = -\sum\limits_{v=1;\,v\ne i}^{n} B_{iv}^{(k)} && 1 \le i \le n \\
		&	&& k = 2,3, \ldots \nonumber
	\end{align}
	
	\begin{equation}\label{eqNC}
		T_i=\int_{0}^{1} \prod_{j=1;j\neq i}^{n}\frac{x-\xi_j}{\xi_i-\xi_j}dx
	\end{equation}

	\section{Discretized eigenvalue problem} \label{ApIevp}
	Discretized eigenvalue problem obtained by applying the LaDQM to Equations \eqref{eq28} to \eqref{eq35}.
	
	\textit{\underline{CC microbeam:}}
	\begin{align} 
		&\ddot q_i + \gamma \dot q_i + \sum_{j=0}^{n+1}{A_{ij}^{(4)} q_j} = - {\alpha _1} \left( \sum_{j=0}^{n+1}{A_{ij}^{(2)} u_j} \right)   \left[    \sum_{j=1}^{n}{T_j \left( u_j \sum_{k=0}^{n+1}{A_{jk}^{(2)} q_k} + q_j \sum_{k=0}^{n+1}{A_{jk}^{(2)} u_k}\right) } \right] \label{eq36} \\
		& \hspace{1.3cm} + \left( \sum_{j=0}^{n+1}{A_{ij}^{(2)} q_j} \right)   \left[ {{N} - {\alpha _1}  \sum_{j=1}^{n}{T_j u_j \left( \sum_{k=0}^{n+1}{A_{jk}^{(2)} u_k}\right) }  } \right]   \nonumber \\
		& \hspace{1.3cm} + {\alpha _2}\frac{{{{ {{V_{DC}}}}^2}}}{{{{\left( {1 - u_i} \right)}^3}}}\left( {2 + \alpha_3 (1 - u_i)} \right) q_i - {\alpha _4} \sum_{j=1}^{n}{T_j p_{ij} } \nonumber  \\ 
		& \hspace{1.3cm} i=2 \cdots n-1  \nonumber \\
		& (1-u_i)^3 \left(  \sum_{k=1}^{n}{B_{ik}^{(2)} p_{kj}}\right)  -3 (1-u_i)^2 \left( \sum_{k=0}^{n+1}{A_{ik}^{(1)} u_k}\right) \left(  \sum_{k=1}^{n}{B_{ik}^{(1)} p_{kj}}\right) \label{eq37} \\
		& \hspace{1.3cm} + \alpha_5 (1-u_i)^3  \left(  \sum_{k=1}^{n}{B_{jk}^{(2)} p_{ik}}\right) = \sigma \left( (1-u_i) \dot p_{ij} - \dot{q}_i\right)  \nonumber\\
		& \hspace{1.3cm} i=2 \cdots n-1 \; \mbox{ ; }\; j=2 \cdots \frac{n+1}{2} \nonumber 
	\end{align}
	subjected to the nondimensional boundary conditions
	\begin{align}
		&q_1 =0,\;  q_n = 0,\; \sum_{j=0}^{n+1}{A_{1j}^{(1)} q_j} =0  \mbox{ and } \sum_{j=0}^{n+1}{A_{nj}^{(1)} q_j} = 0 \label{eq38} \\
		&p_{i1} =0  \mbox{ and } p_{in} =0 \; \mbox{ while } i=1 \cdots n \label{eq39} \\
		& \sum_{k=1}^{n}{B_{1k}^{(1)} p_{kj}} =0 \mbox{ and } \sum_{k=1}^{n}{B_{nk}^{(1)} p_{kj}} = 0 \; \mbox{ while } j=1 \cdots n \nonumber
	\end{align}
	
	\textit{\underline{CF microbeam:}}
	\begin{align} 
		&\ddot q_i + \gamma \dot q_i + \sum_{j=0}^{n+1}{A_{ij}^{(4)} q_j} = {\alpha _2}\frac{{{{ {{V_{DC}}} }^2}}}{{{{\left( {1 - u_i} \right)}^3}}}\left( {2 + \alpha_3 (1 - u_i)} \right)q_i - {\alpha _4} \sum_{j=1}^{n}{T_j p_{ij} }  \label{eq40} \\
		& \hspace{1.3cm} i=2 \cdots n-1  \nonumber \\
		& (1-u_i)^3 \left(  \sum_{k=1}^{n}{B_{ik}^{(2)} p_{kj}}\right)  -3 (1-u_i)^2 \left( \sum_{k=0}^{n+1}{A_{ik}^{(1)} u_k}\right) \left(  \sum_{k=1}^{n}{B_{ik}^{(1)} p_{kj}}\right) \label{eq41} \\
		& \hspace{1.3cm} + \alpha_5 (1-u_i)^3  \left(  \sum_{k=1}^{n}{B_{jk}^{(2)} p_{ik}}\right) = \sigma \left( (1-u_i) \dot p_{ij} - \dot{q}_i\right)  \nonumber\\
		& \hspace{1.3cm} i=2 \cdots n-1 \; \mbox{ ; }\; j=2 \cdots \frac{n+1}{2} \nonumber 
	\end{align}
	subjected to the nondimensional boundary conditions
	\begin{align}
		&q_1 =0,\;  \sum_{j=0}^{n+1}{A_{1j}^{(1)} q_j} =0,\; \sum_{j=0}^{n+1}{A_{nj}^{(2)} q_j} =0  \mbox{ and } \sum_{j=0}^{n+1}{A_{nj}^{(3)} q_j} = 0 \label{eq42} \\
		&p_{i1} =0  \mbox{ and } p_{in} =0 \; \mbox{ while } i=1 \cdots n \label{eq43} \\
		& \sum_{k=1}^{n}{B_{1k}^{(1)} p_{kj}} =0 \mbox{ and } p_{nj} = 0 \; \mbox{ while } j=1 \cdots n \nonumber 
	\end{align}
	
	\section{Transfer and condensed derivative operators} \label{ApII}
	The pressure values in ${\bf z^2}$ are ordered with the node $x_i$, $i = 2, \dots, n-1$, combined with the transverse nodes $j = 2, \dots, \frac{n+1}{2}$. With $n_y = (n-1)/2$ retained nodes $y_j$ for each $x_i$, the operator
	\begin{equation}
		{\bf E} = {\bf I}_{n_w} \otimes {\bf 1}_{n_y},
	\end{equation}
	of dimension $n_p \times n_w$, maps a quantity defined at the nodes $x_i$ of the microbeam onto the nodes $(x_i, y_j)$ of the film. Reflecting the fact that the deflection and the gap are uniform across the width. Here $\otimes$ denotes the Kronecker product. 
	
	Similarly, the operator
	\begin{equation}
		{\bf G} = {\bf I}_{n_w} \otimes {\bf g}^T,
		\qquad
		g_j =
		\begin{cases}
			T_j + T_{n+1-j}, & j = 2, \dots, \dfrac{n-1}{2},\\[4pt]
			T_{\frac{n+1}{2}}, & j = \dfrac{n+1}{2},
		\end{cases}
	\end{equation}
	of dimension $n_w \times n_p$, evaluates the transverse integral at across the width at each $x_i$, $\left({\bf G}\,{\bf z^2}\right)_i = \sum_{j} g_j\, p_{ij} \approx \int_0^1 p(x_i,y)\,dy$, where $T_j$ are the quadrature weights of Equation \eqref{eqNC}. Since only half of the film is retained, the value at each omitted node $n+1-j$ is replaced by that of its symmetric counterpart $j$.
	
	The differential quadrature matrices in \ref{ApI} are applied to all the nodes of the grid, including the boundary and fictitious nodes eliminated through the boundary conditions and, in the $y$-direction, the nodes omitted by symmetry. In order to apply the DQM approximation to retained nodes and apply the required conditions, we let ${\bf R}$ ($n^2 \times n_p$) be the operator that reconstructs the pressure at all nodes of the film, and ${\bf S}$ ($n_p \times n^2$) the operator that selects the retained nodes. And we let ${\bf R}_w$ ($(n+2) \times n_w$) and ${\bf S}_w$ ($n_w \times (n+2)$) be the corresponding operators for the microbeam nodes. The reduced derivative matrices are then
	\begin{equation}
		{\bf a}^{(k)} = {\bf S}_w {\bf A}^{(k)} {\bf R}_w,
		\qquad
		{\bf b}_x^{(k)} = {\bf S}\left({\bf B}^{(k)} \otimes {\bf I}_n\right){\bf R},
		\qquad
		{\bf b}_y^{(k)} = {\bf S}\left({\bf I}_n \otimes {\bf B}^{(k)}\right){\bf R},
	\end{equation}
	
	Since both directions of the film are discretized on the same grid with the same number of points, ${\bf b}_x^{(k)}$ and ${\bf b}_y^{(k)}$ are built from the same weighting coefficients and differ only in the direction of summation and in the boundary conditions incorporated into ${\bf R}$.

%%%%%%%%%%%%%%%%%%%%%%%%%%%%%%%%%%%%%%%%%%%%%%%%%%%%%%%%%%%%%%%%%%%%%%%%%%%%%%%%%%%%%%%%%%%%%%%%%%%%%%%%%%%%%%%%%%%%%%%%%%%%%

\end{document}